\documentclass[prd,preprint,tightenlines,floatfix,
showpacs,preprintnumbers,nofootinbib,eqsecnum,superscriptaddress]{revtex4}

\usepackage{amsmath,amsfonts,amssymb,amstext,mathrsfs}
\usepackage{times}

\usepackage[dvips]{graphicx}
\usepackage{epsf,float}
\usepackage{revsymb}

\usepackage{dcolumn}
\usepackage{braket}
\usepackage{color,xcolor}
\usepackage{graphicx}
\usepackage{subfigure}
\usepackage{multirow}
\usepackage{tabularx}
\usepackage{pstricks}
\usepackage[section]{placeins}
\usepackage{booktabs}
\usepackage{array}

\usepackage{hyperref}

\def\Pom{I\!\!P}
\def\Reg{I\!\!R}

\begin{document}

\title{Central production of \boldmath{$\rho^{0}$} in \boldmath{$pp$} collisions\\
with single proton diffractive dissociation at the LHC}

\author{Piotr Lebiedowicz}
 \email{Piotr.Lebiedowicz@ifj.edu.pl}
\affiliation{Institute of Nuclear Physics Polish Academy of Sciences, Radzikowskiego 152, PL-31-342 Krak\'ow, Poland}

\author{Otto Nachtmann}
 \email{O.Nachtmann@thphys.uni-heidelberg.de}
\affiliation{Institut f\"ur Theoretische Physik, Universit\"at Heidelberg,
Philosophenweg 16, D-69120 Heidelberg, Germany}

\author{Antoni Szczurek
\footnote{Also at University of Rzesz\'ow, PL-35-959 Rzesz\'ow, Poland.}}
\email{Antoni.Szczurek@ifj.edu.pl}
\affiliation{Institute of Nuclear Physics Polish Academy of Sciences, Radzikowskiego 152, PL-31-342 Krak\'ow, Poland}

\begin{abstract}
We consider the $pp \to pp \rho^{0} \pi^{0}$ 
and $pp \to pn \rho^{0} \pi^{+}$ processes at LHC energies.
Our description is based on the nonperturbative framework 
of tensor pomeron and tensor reggeon exchanges.
We discuss the Drell-Hiida-Deck type mechanism 
with centrally produced $\rho^{0}$ meson 
associated with a very forward/backward $\pi N$ system.
The considered processes constitute an inelastic (non-exclusive) background 
to the $p p \to p p \rho^0$ reaction in the case when 
only the centrally produced $\rho^{0}$ meson 
decaying into $\pi^{+} \pi^{-}$ is measured,
the final state protons are not observed, and only 
rapidity-gap conditions are checked experimentally.
We compare our results for the $\gamma \pi^{+} \to \rho^{0} \pi^{+}$ reaction
with the experimental data obtained by the H1 collaboration at HERA.
We present several differential distributions 
for the $pp \to pn \rho^{0} \pi^{+}$ reaction
and estimate the size of the proton dissociative background
to the exclusive $p p \to p p \rho^0$ process.
The ratio of integrated cross sections 
for the inelastic $p p \to p N \rho^0 \pi$ processes,
where $p N \rho^0 \pi$ stands for $p n \rho^0 \pi^{+}$ plus $p p \rho^0 \pi^{0}$,
to the reference reaction $pp \to \ pp \rho^0$ is of order of (7--10)\%.
We present also the ratios of 
the $\rho^0$ rapidity and transverse momentum distributions
for the inelastic $p p \to p N \rho^0 \pi$ 
versus the elastic $pp \to \ pp \rho^0$ reaction.
Our results may be used to investigate 
the $\gamma \pi \to \rho^{0} \pi$ process at LHC energies.
\end{abstract}


\maketitle

\section{Introduction}
\label{sec:intro}

\begin{figure}[!ht]
\includegraphics[width=0.35\textwidth]{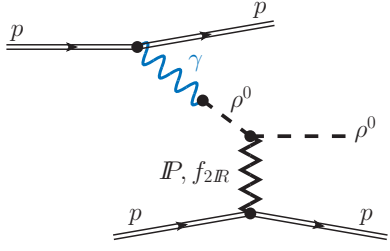}
  \caption{\label{fig:diagrams_pprho0}
  \small
Exclusive central production of $\rho^{0}$ 
in $pp$ collisions by fusion of $\gamma$ and $\Pom$, $f_{2 \Reg}$.
There is also a diagram with the r{\^o}le of the initial protons interchanged.
}
\end{figure}

\begin{figure}[!ht]
(a)\includegraphics[width=0.29\textwidth]{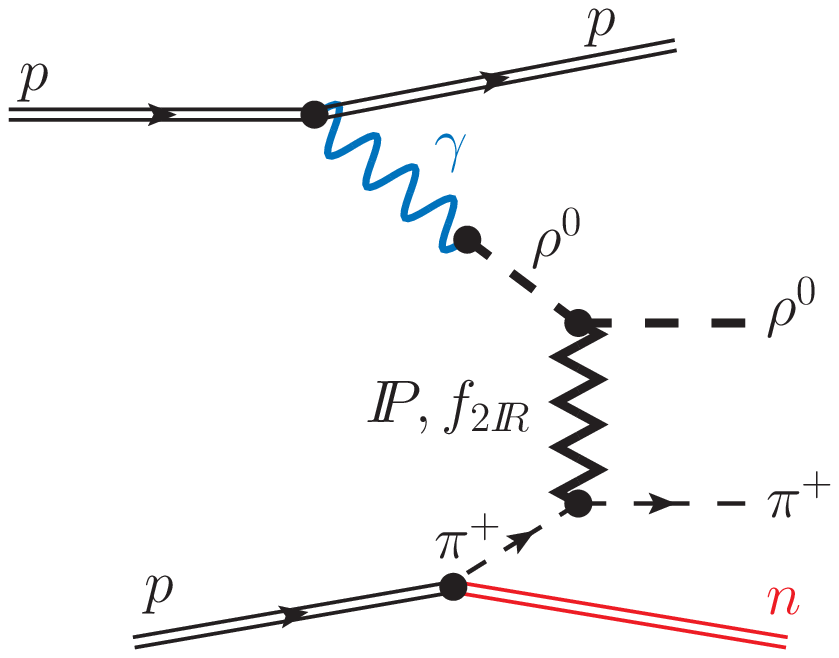}
(b)\includegraphics[width=0.29\textwidth]{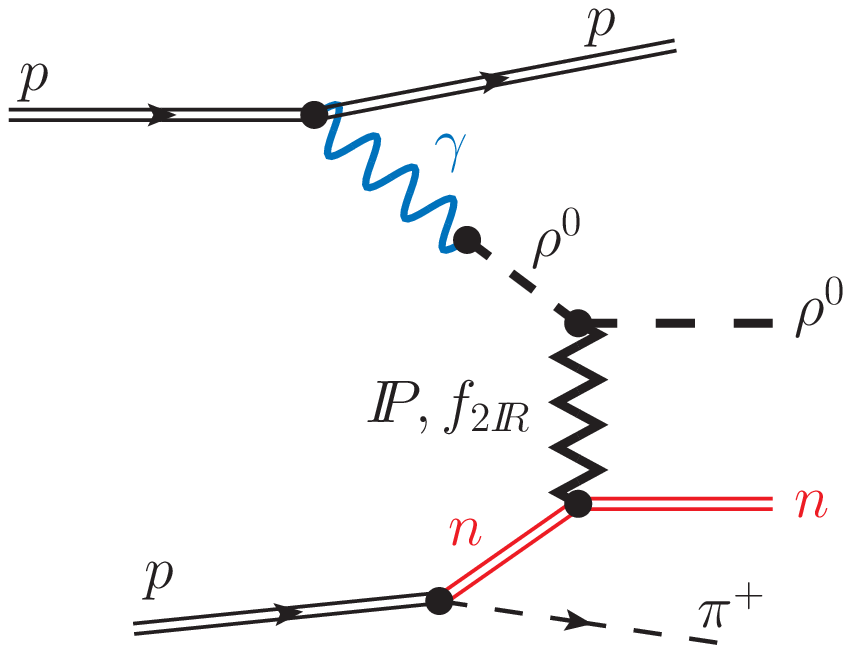}
(c)\includegraphics[width=0.32\textwidth]{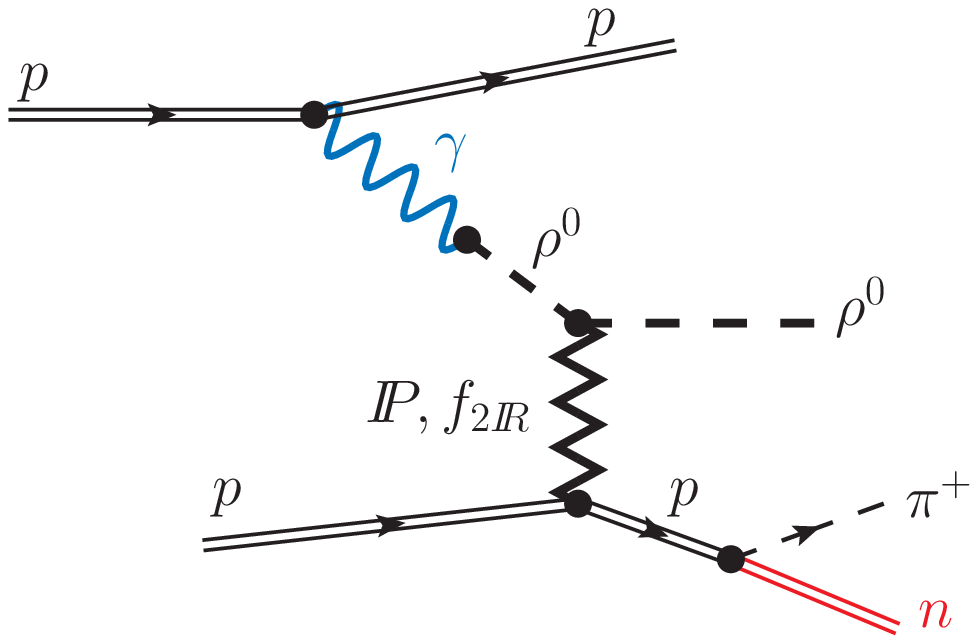}
  \caption{\label{fig:fig0}
  \small
The Born diagrams for processes contributing to 
exclusive $\rho^{0}$ meson photoproduction
associated with a leading neutron in proton-proton collisions.
The diagrams correspond to the Drell-Hiida-Deck type mechanism \cite{Drell:1961zza,Deck:1964hm}
for the pion exchange (a), neutron exchange (b),
and direct production (c).
In the following the incoming proton at the upper side
of the diagrams will be the one with momentum $p_{a}$,
at the lower side with momentum $p_{b}$.
There are also the corresponding diagrams with the r{\^o}le of the initial protons interchanged.
}
\end{figure}

The study of vector meson production in the exclusive $p p \to p p V$ reaction
is one of the important programs for the LHC.
So far the CDF collaboration at Tevatron \cite{Aaltonen:2009kg}
and the LHCb collaboration at the LHC 
\cite{Aaij:2013jxj,Aaij:2014iea,LHCb:2016oce,McNulty:2016sor}
presented their results for ``exclusive'' production of the $J/\psi$ and $\psi'$ mesons.
These vector mesons were observed through their
decay into the $\mu^{+} \mu^{-}$ final state.
Also the cross sections for production of $\Upsilon$ states were measured; see \cite{Aaij:2015kea}.
However, so far forward going protons were not measured at the LHC.
Instead, the LHCb collaboration checks only the rapidity gaps around the measured vector meson. 
Therefore, one is not sure whether the reaction is fully exclusive
or whether there are contributions from dissociation of one or both protons
in the final state. 
In the following we shall consider as reference reaction exclusive central
$\rho^{0}$ production in $pp$ collisions (see Fig.~\ref{fig:diagrams_pprho0})
\begin{eqnarray}
p + p \to p + \rho^{0} + p \,.
\label{2to3_reaction_intro}
\end{eqnarray}
Here the $\rho^{0}$ is produced by the fusion of a virtual photon
emitted from one proton and a pomeron plus $f_{2 \Reg}$ reggeon from the other proton.
In the inelastic case the proton emitting the photon, 
or the one emitting $\Pom$, $f_{2 \Reg}$, or both protons may dissociate.
If these remnants from the dissociated protons have low invariant mass
they constitute a background to the reaction (\ref{2to3_reaction_intro}).
Experimentally this background is notoriously difficult to handle.
Also from the theory point of view these breakup reactions have rarely been studied.
In \cite{Schafer:2016gwq,Cisek:2016kvr} the electromagnetic dissociation 
was estimated to be of the order of 10\%
(for excited states $M_{X} < 2$~GeV)
of the exclusive cross section for $J/\psi$ production.
Here we wish to make first estimates in the case of 
diffractive proton excitation for $\rho^{0}$ production. 
That is, we shall study the case where the proton
at the $\Pom$, $f_{2 \Reg}$ side of Fig.~\ref{fig:diagrams_pprho0} breaks up into
a $\pi^{+} n$ continuum state
\begin{eqnarray}
p + p \to p + \rho^{0} + n + \pi^{+}\,.
\label{2to4_reaction_intro}
\end{eqnarray}
The corresponding diagrams are shown in Fig.~\ref{fig:fig0}.
This gives then a leading neutron on one side of the collision.
Such leading neutrons can be rather easily detected
with special forward detectors \cite{ATLAS:2007aa,Grachov:2008qg}.
The reaction (\ref{2to4_reaction_intro}) can be seen 
as a Drell-Hiida-Deck mechanism \cite{Drell:1961zza,Deck:1964hm}
where a proton is dissociating into the ($n$,~$\pi^{+}$) system
which scatters elastically on the $\rho^{0}$ meson
via the exchange of the pomeron and $f_{2 \Reg}$ reggeon.
In addition to the pion exchange mechanism
(see corresponding diagram of Fig.~\ref{fig:fig0}~(a))
two further contributions (diagrams (b) and (c)) must be included.
The two diagrams (b) and (c) give contributions to the total
scattering amplitude with similar magnitude but opposite sign;
see e.g. \cite{Tsarev:1974nr,Berger:1975qa,
Ponomarev:1975ru,Ponomarev:1976nv,Zotov:1978,Tarasiuk:1979ss}. 
Therefore, in most of the phase space,
in particular, at small momentum transfer squared at the $p \to n$ vertex, 
the contributions of diagrams (b) and (c)
essentially cancel such that the contribution of diagram (a) dominates 
the cross section. 
Here we shall concentrate on the pion exchange mechanism;
see Fig.~\ref{fig:fig0} (a).
For a recent consideration of the Drell-Hiida-Deck mechanism \cite{Drell:1961zza,Deck:1964hm}
at LHC energies, see e.g. the discussion of the $pp \to pp \pi^{0}$
reaction in \cite{Lebiedowicz:2013vya}.

The production of leading neutrons was studied in deep-inelastic 
$ep$ scattering (semi-inclusive $e+p \to e+n+X$ process) 
at HERA \cite{Aaron:2010ab,Andreev:2014zka}.
Very recently the first measurement of exclusive photoproduction of $\rho^{0}$ mesons 
associated with leading neutrons ($\gamma p \to \rho^0 n \pi^+$) 
was presented by the H1 collaboration \cite{H1:2015bxa}.
The HERA experimental results indicate that the production
of neutrons carrying a large fraction of the proton beam energy 
is indeed dominated by the pion exchange process.
The description of these leading neutron processes still is a theoretical challenge.
Exclusive processes with a leading neutron in $ep$ collisions were discussed recently in 
the color dipole approach \cite{Goncalves:2015mbf}
using the flux of virtual pions emitted by the proton.
For related work on the exclusive vector-meson ($\rho$, $\phi$ and $J/\psi$) production
associated with a leading neutron see \cite{Goncalves:2016uhj}.
In~the~following we shall compare our results for the pion exchange mechanism
(see Fig.~\ref{fig:fig0}~(a)) with those obtained in \cite{Goncalves:2016uhj}.

In \cite{Lebiedowicz:2014bea} we considered the reaction $p p \to p p (\rho^0 \to \pi^+ \pi^-)$
within the tensor-pomeron approach formulated in \cite{Ewerz:2013kda}.
In \cite{Ewerz:2016onn} three models for the soft pomeron,
tensor, vector, and scalar, were compared
with the STAR experimental data on polarised high-energy $pp$ scattering \cite{Adamczyk:2012kn}.
Only the tensor-pomeron model was found to be consistent
with the general rules of quantum field theory and the data from \cite{Adamczyk:2012kn}.
Recently, both the $\rho^{0}$-photoproduction 
and the purely diffractive contributions have been discussed
for the $p p \to p p \pi^+ \pi^-$ reaction; see \cite{Lebiedowicz:2016ioh}.
In the present paper we wish to make first predictions 
for the process (\ref{2to4_reaction_intro}) within the same framework.

Motivated by the study of two of us of diffractive $\pi^{0}$-strahlung 
production \cite{Lebiedowicz:2013vya} we consider here only the contributions 
related to a $p \to \pi^+ n$ transition
which is interesting by itself 
(the $p \to \pi^0 p$ transition can be done analogously).
A related hadronic bremsstrahlung mechanism of difractive production
of $\omega N$ states has been discussed in \cite{Cisek:2011vt}.
In general, there are also contributions due to
diffractive excitation of resonances, $N^{*}$ states,
and their subsequent decays into the $\pi N$ channel.
For an analysis of proton diffractive dissociation 
to $N^{*}$ states see~\cite{Jenkovszky:2010ym,Jenkovszky:2012hf}.

Our paper is organized as follows. In Sec.~\ref{sec:2to2}
we present the basic formulae for the $\gamma \pi^{+} \to \rho^{0} \pi^{+}$ reaction
within the tensor-pomeron approach
and compare our results with the H1 experimental data.
In Sec.~\ref{sec:2to4} we consider the $pp \to pn \rho^{0} \pi^{+}$ process
shown in Fig.~\ref{fig:fig0}~(a).
Sec.~\ref{sec:results} contains numerical results
for total and differential cross sections calculated for the LHC energies.
We~present also the ratios of 
the $\rho^0$ rapidity and transverse momentum distributions
for the inelastic $p p \to p N \rho^0 \pi$ processes,
where $p N \rho^0 \pi$ stands for $p n \rho^0 \pi^{+}$ plus $p p \rho^0 \pi^{0}$,
versus the elastic reaction $pp \to \ pp \rho^0$.
Sec.~\ref{sec:concl} presents our conclusions.

\section{The reaction $\gamma \pi^{+} \to \rho^{0} \pi^{+}$}
\label{sec:2to2}

\begin{figure}[!ht]
\includegraphics[width=0.3\textwidth]{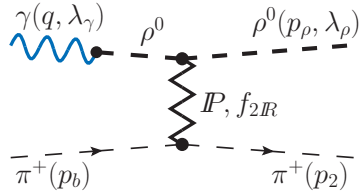}
  \caption{\label{fig:diagram_2to2}
  \small
Diagram for $\gamma \pi^{+} \to \rho^{0} \pi^{+}$ with pomeron and $f_{2 \Reg}$ reggeon exchange.
We use the vector meson dominance model and the corresponding relation for the $\gamma$-$\rho^{0}$ coupling.
}
\end{figure}

As a first ingredient for our calculations we consider the reaction 
(see Fig.~\ref{fig:diagram_2to2})
\begin{eqnarray}
\gamma(q,\lambda_{\gamma}) + \pi^{+}(p_{b}) \to \rho(p_{\rho},\lambda_{\rho}) + \pi^{+}(p_{2})
\label{2to2_reaction}
\end{eqnarray}
for real photons. 
Here the four-momenta and the helicities, $\lambda_{\gamma} = \pm 1$
and $\lambda_{\rho} = \pm 1, 0$, are indicated in brackets.
We use standard kinematic variables
\begin{eqnarray}
&&s = W_{\gamma \pi}^{2} = (p_{b} + q)^{2} = (p_{2} + p_{\rho})^{2}\,, \nonumber \\
&&t = (p_{2} - p_{b})^{2} = (p_{\rho} -q)^{2}\,.
\label{2to2_kinematic}
\end{eqnarray}
%

The differential cross section for the reaction (\ref{2to2_reaction}) 
for unpolarised photons and no observation of the $\rho^{0}$ polarisation
is given by
\begin{eqnarray}
\frac{d\sigma}{dt} =
\frac{1}{16 \pi (s- m_{\pi}^{2})^{2}} \;
\frac{1}{2} \sum_{\lambda_{\gamma}, \lambda_{\rho}}
|{\cal M}_{\lambda_{\gamma} \pi^{+} \to \lambda_{\rho} \pi^{+}}|^{2} \,.
\label{xsecttion_diff_2to2}
\end{eqnarray}
The ${\cal T}$-matrix element is
\begin{eqnarray}
{\cal M}_{\lambda_{\gamma} \pi^{+} \to \lambda_{\rho} \pi^{+}}=
\Braket{\rho^{0}(p_{\rho},\lambda_{\rho}),\pi^{+}(p_{2})
|{\cal T}|
\gamma(q,\lambda_{\gamma}),\pi^{+}(p_{b})}\,.
\label{2to2_Tmatrix}
\end{eqnarray}

The amplitude via the tensor-pomeron exchange is written as
\begin{eqnarray}
&&{\cal M}^{(\Pom)}_{\lambda_{\gamma} \pi^{+} \to \lambda_{\rho} \pi^{+}} =
(-i) \, \left( \epsilon^{(\rho)\,\mu}(p_{\rho},\lambda_{\rho})\right)^* \,
i\Gamma_{\mu \nu \alpha \beta}^{(\Pom \rho \rho)}(p_{\rho},q) \,  
i\Delta^{(\rho)\,\nu \kappa}(q) \,
i\Gamma^{(\gamma \to \rho)}_{\kappa \sigma}(q)\,
\epsilon^{(\gamma)\,\sigma} (q,\lambda_{\gamma})
\nonumber \\ 
&& \qquad \qquad \qquad \quad \times i\Delta^{(\Pom)\,\alpha \beta, \delta \eta}(s,t)\, 
i\Gamma_{\delta \eta}^{(\Pom \pi \pi)}(p_{2},p_{b}) \,,
\label{rhop_tot_opt_aux0}
\end{eqnarray}
where $\epsilon^{(\gamma)}$ and $\epsilon^{(\rho)}$ are the polarisation vectors 
for photon and $\rho^{0}$ meson, respectively.

The $\Pom \rho \rho$ vertex is given in \cite{Ewerz:2013kda} by formula (3.47).

The effective propagator of the tensor-pomeron exchange is written as
(see (3.10) of \cite{Ewerz:2013kda}):
\begin{eqnarray}
i \Delta^{(\Pom)}_{\mu \nu, \kappa \lambda}(s,t) = 
\frac{1}{4s} \left( g_{\mu \kappa} g_{\nu \lambda} 
                  + g_{\mu \lambda} g_{\nu \kappa}
                  - \frac{1}{2} g_{\mu \nu} g_{\kappa \lambda} \right)
(-i s \alpha'_{\Pom})^{\alpha_{\Pom}(t)-1}
\label{pomeron_propagator}
\end{eqnarray}
and fulfils the following relations
\begin{eqnarray}
&&\Delta^{(\Pom)}_{\mu \nu, \kappa \lambda}(s,t) = 
\Delta^{(\Pom)}_{\nu \mu, \kappa \lambda}(s,t) =
\Delta^{(\Pom)}_{\mu \nu, \lambda \kappa}(s,t) =
\Delta^{(\Pom)}_{\kappa \lambda, \mu \nu}(s,t) \,,\nonumber \\
&&g^{\mu \nu} \Delta^{(\Pom)}_{\mu \nu, \kappa \lambda}(s,t) = 0, \quad 
g^{\kappa \lambda} \Delta^{(\Pom)}_{\mu \nu, \kappa \lambda}(s,t) = 0 \,.
\label{pomeron_propagator_aux}
\end{eqnarray}
Here the pomeron trajectory $\alpha_{I\!\!P}(t)$ is assumed to be 
of standard linear form with intercept slightly above 1:
\begin{eqnarray}
\alpha_{\Pom}(t) = \alpha_{\Pom}(0)+\alpha'_{\Pom}\,t, \quad 
\alpha_{\Pom}(0) = 1.0808,\quad \alpha'_{\Pom} = 0.25 \; \mathrm{GeV}^{-2}\,.
\label{pomeron_trajectory}
\end{eqnarray}

For the $\Pom \pi \pi$ vertex we have (see Eq.~(3.45) of \cite{Ewerz:2013kda}
and (B.69) of \cite{Bolz:2014mya})
\begin{eqnarray}
i\Gamma_{\mu \nu}^{(\Pom \pi \pi)}(k',k)=
-i 2 \beta_{\Pom \pi \pi} 
\left[ (k'+k)_{\mu}(k'+k)_{\nu} - \frac{1}{4} g_{\mu \nu} (k' + k)^{2} \right] \, F_{M}((k'-k)^2)\,.
\label{vertex_pompipi}
\end{eqnarray}
Here $\beta_{\Pom \pi \pi} = 1.76$~GeV$^{-1}$ and
$F_{M}(t)$ is the pion electromagnetic form factor 
in a parametrization valid for $t < 0$,
\begin{eqnarray}
F_{M}(t)=\frac{1}{1-t/\Lambda_{0}^{2}}\,,
\label{F_pion}
\end{eqnarray}
where $\Lambda_{0}^{2} = 0.5$~GeV$^{2}$; 
see e.g.~(3.22) of \cite{Donnachie:2002en} and (3.34) of \cite{Ewerz:2013kda}.

Including $f_{2 \Reg}$ exchange we obtain for the amplitude (\ref{2to2_Tmatrix})
\begin{eqnarray}
{\cal M}^{(\Pom + f_{2 \Reg})}_{\lambda_{\gamma} \pi^{+} \to \lambda_{\rho} \pi^{+}} (s, t) =
&&i e \dfrac{m_{\rho}^{2}}{\gamma_{\rho}}\,
\Delta_{T}^{(\rho)}(0) \,
(\epsilon^{(\rho)\, \mu}(p_{\rho},\lambda_{\rho}))^*
\epsilon^{(\gamma)\, \nu}(q,\lambda_{\gamma}) 
V_{\mu \nu \kappa \lambda}(s,t,q,p_{\rho})
\nonumber \\
&& \times
2 (p_2+p_b)^{\kappa} (p_2+p_b)^{\lambda} \, [F_{M}(t)]^{2} \,.
\label{rhop_tot_opt_aux}
\end{eqnarray}
Here we use (3.2), (3.10), (3.12), (3.23),
(3.45), (3.47), (3.53), and (3.55) of \cite{Ewerz:2013kda},
in particular, we have $M_{0} \equiv 1$~GeV,
$4 \pi/ \gamma_{\rho}^{2} =  0.496$,
$(\Delta_{T}^{(\rho)}(0))^{-1} = -m_{\rho}^{2}$.
The function $V_{\mu \nu \kappa \lambda}(s,t,q,p_{\rho})$ has the form
\begin{eqnarray}
&&V_{\mu \nu \kappa \lambda}(s,t,q,p_{\rho})=\frac{1}{4s}
\nonumber \\
&& \times \bigg\{
2 \Gamma_{\mu \nu \kappa \lambda}^{(0)}(p_{\rho},-q)
\left[
2 \beta_{\Pom \pi\pi} \, a_{\Pom \rho\rho} 
(- i s \alpha'_{\Pom})^{\alpha_{\Pom}(t)-1}
+ (2 M_0)^{-1} g_{f_{2 \Reg} \pi \pi} \, a_{f_{2 \Reg} \rho \rho}
(- i s \alpha'_{\Reg_{+}})^{\alpha_{\Reg_{+}}(t) -1} 
\right]
\nonumber \\
&& -\Gamma_{\mu \nu \kappa \lambda}^{(2)}(p_{\rho},-q)
\left[
2 \beta_{\Pom \pi\pi} \, b_{\Pom \rho \rho} 
(- i s \alpha'_{\Pom} )^{\alpha_{\Pom}(t)-1} 
+ (2 M_0)^{-1} g_{f_{2 \Reg} \pi \pi} \, b_{f_{2 \Reg} \rho \rho}
(- i s \alpha'_{\Reg_{+}})^{\alpha_{\Reg_{+}}(t) -1} 
\right] \bigg\}\,.\nonumber \\
\label{rhop_tot_opt_aux2}
\end{eqnarray}
The explicit tensorial functions 
$\Gamma_{\mu \nu \kappa \lambda}^{(i)}(p_{\rho},-q)$, 
$i$ = 0, 2, are given in \cite{Ewerz:2013kda}, 
formulae (3.18) and (3.19), respectively.
We take the parameters occurring in (2.12) from \cite{Ewerz:2013kda}
and for the coupling constants $a$ and $b$ the set~A
given by (2.15) of \cite{Lebiedowicz:2014bea}.
In this way we described the experimental data
for elastic photoproduction of $\rho^{0}$ meson in the $\gamma p \to \rho^{0} p$ reaction 
fairly well for energies $W_{\gamma p} \gtrsim 8$~GeV; see Fig.~4 (left panel) of \cite{Lebiedowicz:2014bea}. 

\begin{figure}[!ht]
(a)\includegraphics[width=0.44\textwidth]{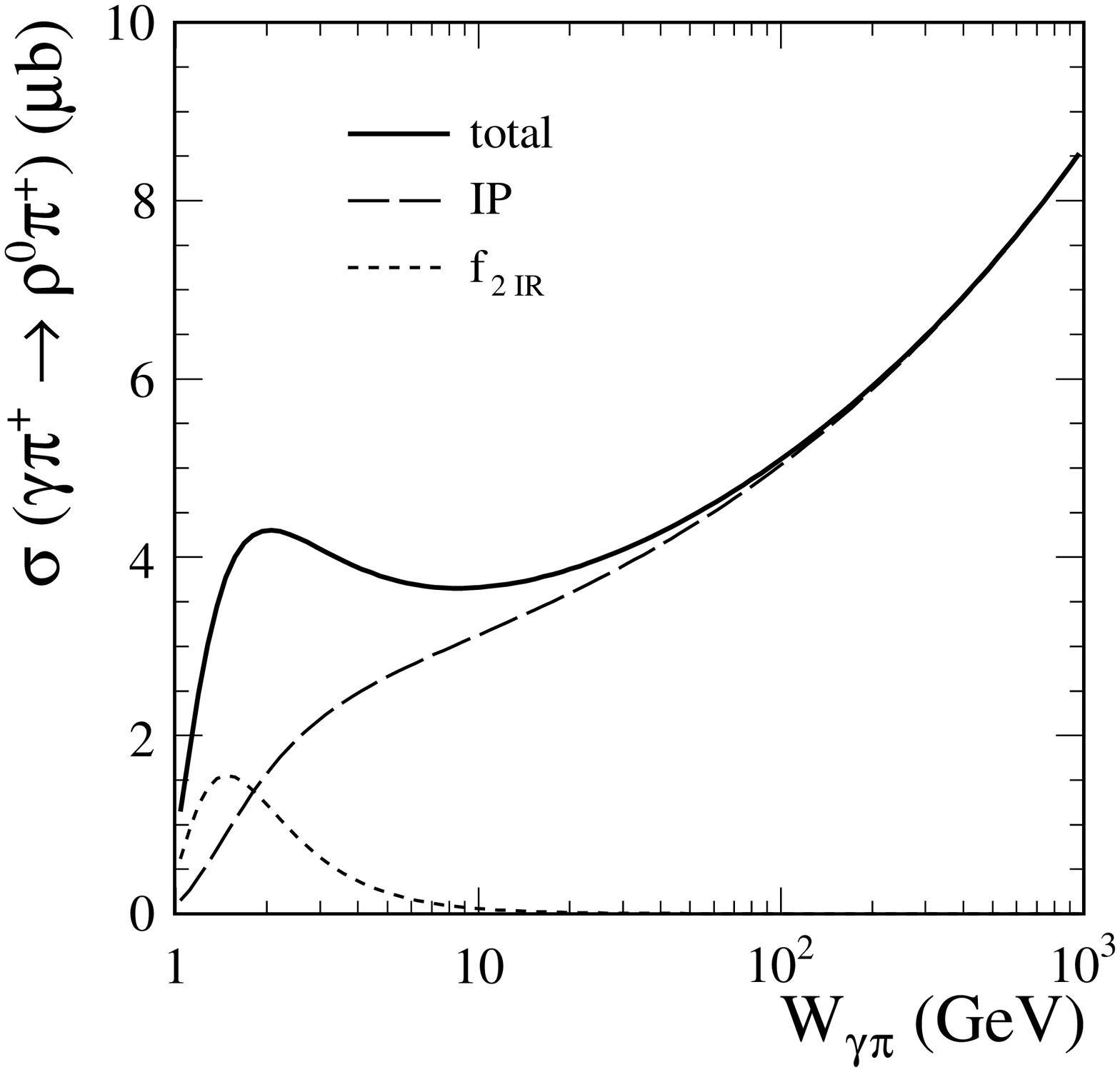}
(b)\includegraphics[width=0.44\textwidth]{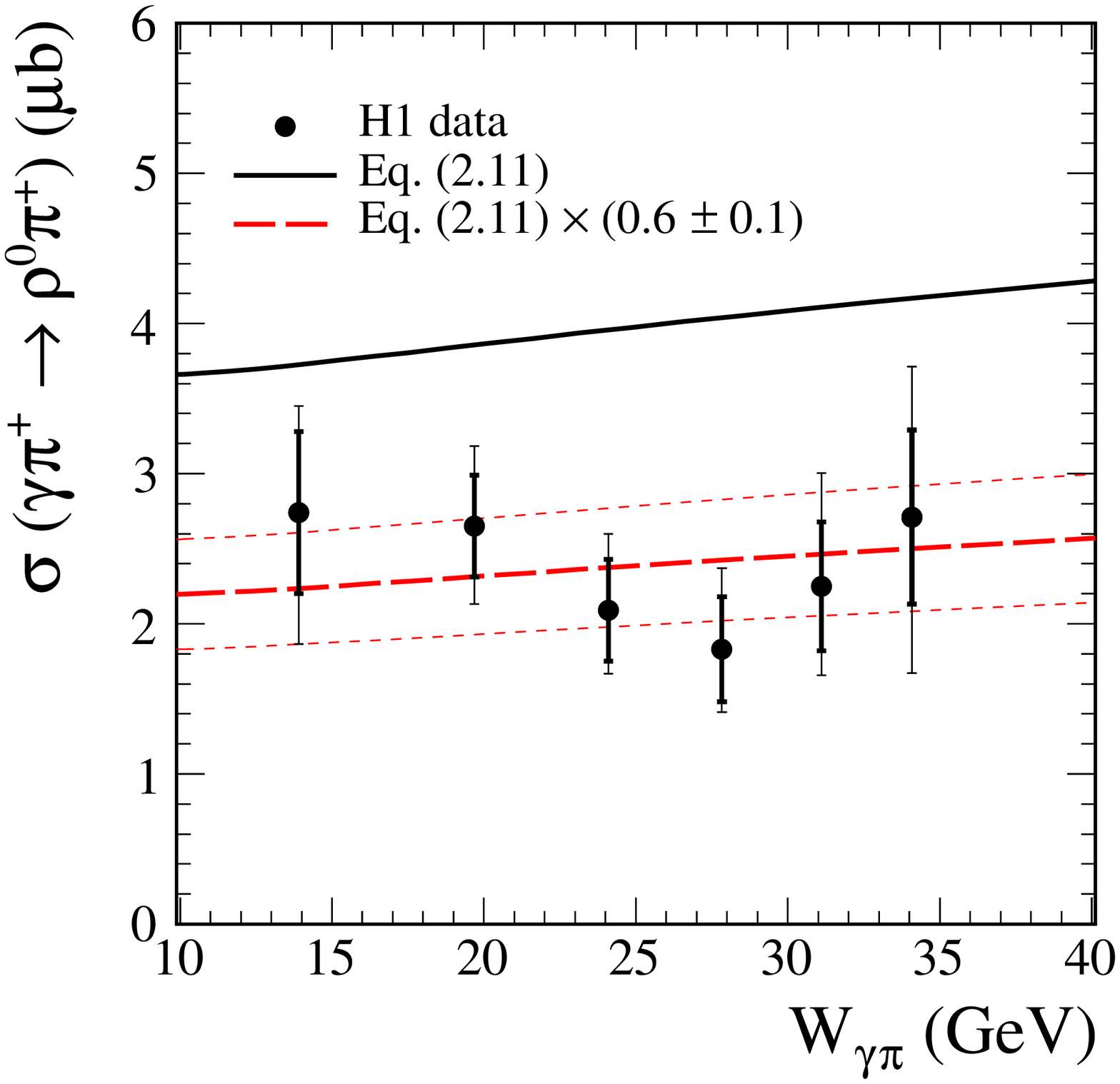}
(c)\includegraphics[width=0.44\textwidth]{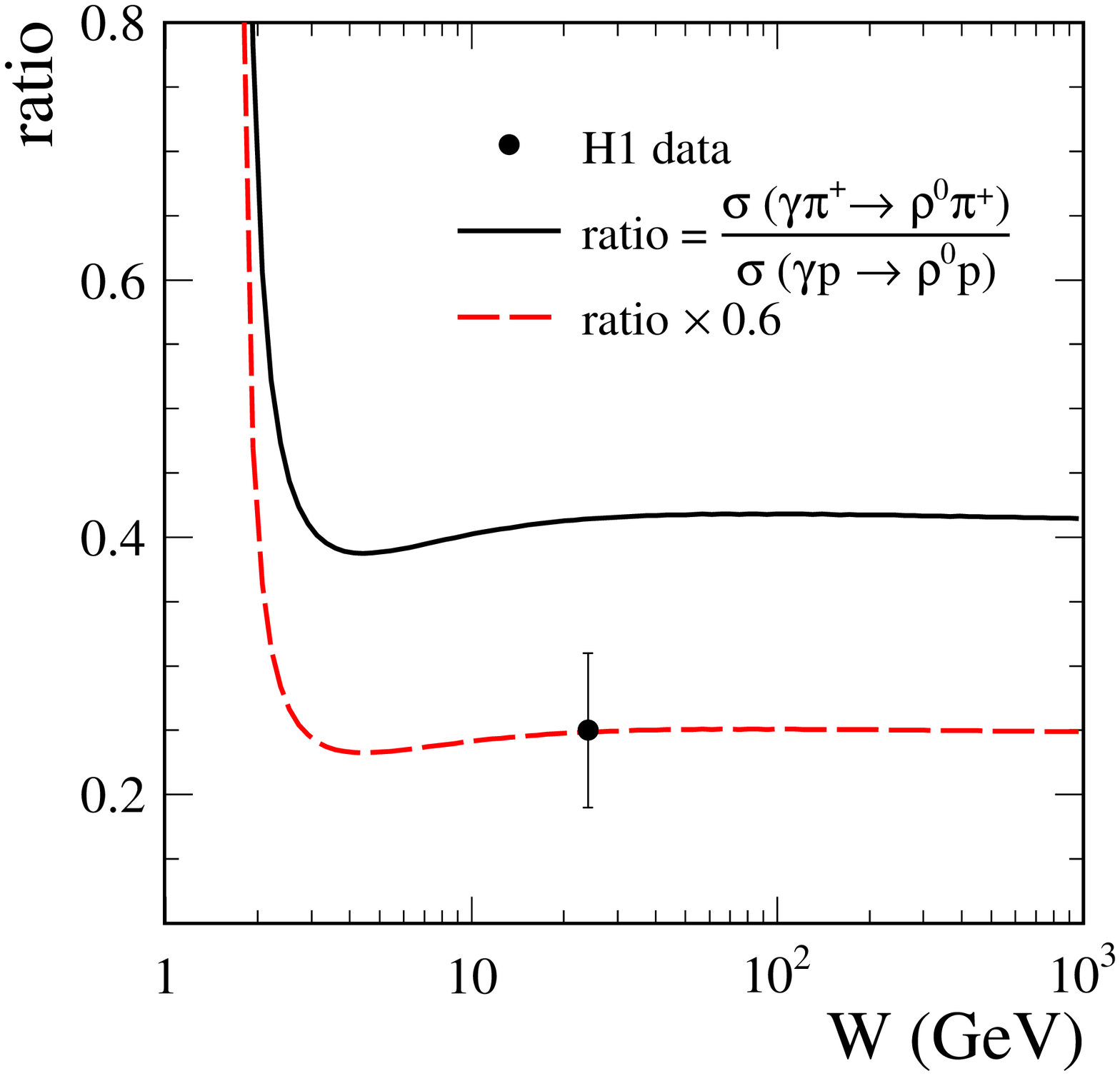}
\caption{\label{fig:photoprod_rho2}
\small
(a): Cross section of elastic $\rho^{0}$ photoproduction on the pion
as a function of the center-of-mass energy $W_{\gamma \pi}$.
The solid line corresponds to results with both the tensor pomeron and tensor $f_{2 \Reg}$ exchanges.
The individual pomeron and reggeon exchange contributions
denoted by the long-dashed and short-dashed lines, respectively, are presented.
(b): Extended view of the $W_{\gamma \pi}$ region where data from \cite{H1:2015bxa} exist.
We show the result from (\ref{rhop_tot_opt_aux})
and the result multiplied by the factor $0.6 \pm 0.1$. 
The experimental data are from Table~9 of \cite{H1:2015bxa}.
The inner error bars represent the total experimental uncertainty,
the outer errors are experimental and model uncertainties are added in quadrature.
(c): The ratio of cross sections $\sigma(\gamma \pi^{+} \to \rho^{0} \pi^{+})$ and
$\sigma(\gamma p \to \rho^{0} p)$. The data point at $W = 24$~GeV is taken from \cite{H1:2015bxa}.
}
\end{figure}

Fig.~\ref{fig:photoprod_rho2}~(a) shows 
the integrated cross section for the $\gamma \pi^{+} \to \rho^{0} \pi^{+}$ reaction
as a function of the center-of-mass energy.
Our result is compared with the H1 experimental results \cite{H1:2015bxa}
given in Table~9 of \cite{H1:2015bxa}
in the $W_{\gamma \pi}$ region where these data exist, see Fig.~\ref{fig:photoprod_rho2}~(b).
The experimental photon-pion cross sections were extracted from
the exclusive $\rho^{0}$ production associated with a leading neutron,
$\gamma p \to \rho^{0} n \pi^{+}$,
using the differential cross section $d\sigma_{\gamma p}/dx_{L}$
and the pion flux integrated over the range $p_{t,n} < 0.2$~GeV; see \cite{H1:2015bxa}.
We see from Fig.~\ref{fig:photoprod_rho2}~(b) that the $W_{\gamma \pi}$ shape
of the data is rather well represented by our results.
Note that for this it is important to have both contributions,
pomeron and $f_{2 \Reg}$ reggeon exchange.
But to obtain the normalization of the data we have to multiply our results with a factor $K = 0.6$.
The same message can be determined from Fig.~\ref{fig:photoprod_rho2}~(c)
where we show the ratio of cross sections for elastic $\rho^{0}$ photoproduction
on pions and protons.
To calculate the cross section for the $\gamma p \to \rho^{0} p$ process
we use formula (2.1) given in \cite{Lebiedowicz:2014bea}.
We can see that, according to our calculation,
the ratio $\sigma^{\gamma \pi}/\sigma^{\gamma p}$
is slightly above $0.4$ for $W > 20$~GeV.
However, in \cite{H1:2015bxa} a significantly smaller value of the ratio, 
$0.25 \pm 0.06$ (at the corresponding HERA energy $<W> = 24$~GeV), 
has been obtained.
This inconsistency for the normalization (Fig.~\ref{fig:photoprod_rho2}~(b)) 
and for the ratio shown in Fig.~\ref{fig:photoprod_rho2}~(c) may come from 
the influence of rescattering (absorptive) corrections
which may be essential for the exclusive reaction $\gamma p \to \rho^{0} n \pi^{+}$
studied experimentally \cite{H1:2015bxa}.
The absorption factor estimated in \cite{H1:2015bxa} is $K_{abs} = 0.44 \pm 0.11$
whereas we find here $K = 0.6 \pm 0.1$.

\section{The reaction $pp \to pn \rho^{0} \pi^{+}$}
\label{sec:2to4}

Here we discuss the exclusive production 
of $\rho^{0}$ meson associated with a forward $\pi^{+} n$ system 
in proton-proton collisions,
\begin{eqnarray}
p(p_{a},\lambda_{a}) + p(p_{b},\lambda_{b}) \to
p(p_{1},\lambda_{1}) + \rho(p_{\rho},\lambda_{\rho}) + \pi^{+}(p_{\pi}) + n(p_{2},\lambda_{2}) \,.
\label{2to4_reaction}
\end{eqnarray}
The kinematic variables for (\ref{2to4_reaction}) are
\begin{eqnarray}
&&q_{1} = p_{a} - p_{1}\,, \quad q_{2} = p_{b} - p_{2}\,, \nonumber \\
&&t_{1} = q_{1}^{2}\,, \quad t_{2} = q_{2}^{2}\,, \nonumber \\
&&\hat{s} = (q_{1} + q_{2})^{2} = (p_{\rho} + p_{\pi})^{2}\,, \nonumber\\
&&\hat{t} = (q_{1} - p_{\rho})^{2} = (q_{2} - p_{\pi})^{2}\,. 
\label{2to4_kinematic}
\end{eqnarray}

The ``bare'' amplitude (excluding rescattering effects) 
for the $\gamma \Pom$ exchange, see diagram~(a) in~Fig.~\ref{fig:fig0},
can be written as follows:
\begin{eqnarray}
&&{\cal M}^{(\gamma \Pom)}_{\lambda_{a} \lambda_{b} \to \lambda_{1} \lambda_{2} \lambda_{\rho}}
= (-i)
\bar{u}(p_{1}, \lambda_{1}) 
i\Gamma^{(\gamma pp)}_{\mu}(p_{1},p_{a}) 
u(p_{a}, \lambda_{a}) \nonumber \\
&&\qquad \times  
i\Delta^{(\gamma)\,\mu \sigma}(q_{1})\, 
i\Gamma^{(\gamma \to \rho)}_{\sigma \nu}(q_{1})\,
i\Delta^{(\rho)\,\nu \rho_{1}}(q_{1}) \,
\left( \epsilon^{(\rho)\,\rho_{2}}(p_{\rho}, \lambda_{\rho}) \right)^*
\nonumber \\
&& \qquad \times 
i\Gamma^{(\Pom \rho \rho)}_{\rho_{2} \rho_{1} \alpha \beta}(p_{\rho},q_{1})\, 
i\Delta^{(\Pom)\,\alpha \beta, \delta \eta}(\hat{s}, \hat{t})
\nonumber \\
&& \qquad \times 
i\Gamma^{(\Pom \pi \pi)}_{\delta \eta}(p_{\pi},q_{2})\,
i\Delta^{(\pi)}(t_{2}) \,
\bar{u}(p_{2}, \lambda_{2}) 
i\Gamma^{(\pi p n)}(p_{2},p_{b}) 
u(p_{b}, \lambda_{b}) \,.
\label{amplitude_gamma_pomeron_2to4}
\end{eqnarray}
The $\gamma pp$ vertex is given in \cite{Ewerz:2013kda} by formula (3.26).
For the pion-nucleon vertex we have
\begin{eqnarray}
i\Gamma^{(\pi pp)}(p',p) = \frac{1}{\sqrt{2}}\, i\Gamma^{(\pi pn)}(p',p) =
-\gamma_{5}\,g_{\pi NN} \,F_{\pi NN}\bigl((p'-p)^2\bigr)\,.
\label{vertex_piNN}
\end{eqnarray}
The general expressions for the pion-nucleon coupling are given in \cite{Dumbrajs:1983jd}.
We have for the $\pi^{0}pp$ coupling constant $g_{\pi NN} >0$
and $g_{\pi NN}^{2}/(4 \pi) = 14.4$
as a typical value quoted in the literature;
see for instance \cite{Dumbrajs:1983jd,Ericson:2000md}.
The form factor $F_{\pi NN}(t)$ 
is normalized to unity at the on-shell point $F_{\pi NN}(m_{\pi}^{2}) = 1$
and parametrised here as
\begin{eqnarray}
F_{\pi NN}(t) = \exp \left(\frac{t-m_{\pi}^{2}}{\Lambda^{2}} \right)\,,
\label{ff_piNN}
\end{eqnarray}
where $\Lambda$ could be adjusted to experimental data. 
We take $\Lambda = 1$~GeV and 1.2~GeV for comparison.

In the high-energy small-angle approximation we get,
including also $\gamma f_{2 \Reg}$ exchange,
\begin{eqnarray}
&& {\cal M}^{(\gamma \Pom + \gamma f_{2 \Reg})}_{\lambda_{a}\lambda_{b}\to\lambda_{1}\lambda_{2}\lambda_{\rho}}
\simeq - e^{2} \frac{m_{\rho}^{2}}{\gamma_{\rho}} (p_1 + p_a)^{\rho_{1}} \delta_{\lambda_{1}\lambda_{a}} F_1(t_1) 
\nonumber \\
&&\qquad \times  
\frac{1}{t_{1}}\,
\Delta^{(\rho)}_{T}(t_{1}) \, \left( \epsilon^{(\rho)\,\rho_{2}}(p_{\rho},\lambda_{\rho})\right)^* \,
\tilde{F}^{(\rho)}(t_{1}) \,
V_{\rho_{2} \rho_{1} \alpha \beta}(\hat{s}, \hat{t}, q_{1}, p_{\rho}) 
[F_{M}(\hat{t})]^{2}
\nonumber \\
&& \qquad \times 
2 (p_{\pi} + q_{2})^{\alpha} (p_{\pi} + q_{2})^{\beta}
\frac{F_{\pi NN}(t_2) \hat{F}_{\pi}(t_2)}{t_{2}-m_{\pi}^{2}}
\sqrt{2}g_{\pi NN}\, \bar{u}(p_{2}, \lambda_{2}) 
\gamma_{5}
u(p_{b}, \lambda_{b}) \,.
\label{amplitude_approx_2to4}
\end{eqnarray}
Here we take only the Dirac form factor of the proton $F_{1}(t_{1})$; see (3.29) of \cite{Ewerz:2013kda}.
We have also included form factors $\tilde{F}^{(\rho)}(t_{1})$ and $\hat{F}_{\pi}(t_2)$
taking into account that the $\rho$ created in the $\gamma$-$\rho$ transition 
and the $\pi^{+}$ emitted from the proton $p(p_{b})$,
respectively, are off shell. 
We assume in the calculations presented in this paper
that $\hat{F}_{\pi}(t) = F_{\pi NN}(t)$
which corresponds to the exponential form for $\hat{F}_{\pi}$; 
see (3.17) of \cite{Lebiedowicz:2016ioh}.
For $\tilde{F}^{(\rho)}(t)$ we take a form as given in (B.85) of \cite{Bolz:2014mya}
with $\Lambda_{\rho} = 2$~GeV and $n_{\rho} =0.5$;
see also (3.9) and the discussion of Fig.~8 in \cite{Lebiedowicz:2014bea}.

\section{First results}
\label{sec:results}

Now we show numerical results for the reaction $p p \to p n \rho^0 \pi^+$.
Our preliminary studies here are done in the Born approximation 
(neglecting absorptive corrections).
We get the total cross section from integrating
over the whole phases space: $\sigma =$ 387.2 (463.2)~nb at $\sqrt{s} = 7$ (13)~TeV.
These results are obtained for the diagram (a) of Fig.~\ref{fig:fig0} 
plus the one with the r{\^o}le of the initial protons interchanged,
with both $\Pom$ and $f_{2 \Reg}$ reggeon exchanges
and for $\Lambda = 1$~GeV in (\ref{ff_piNN})~\footnote{
For comparison, for $\sqrt{s} = 13$~TeV, we get the cross section equal to 407.4~nb
when only the $\Pom$ exchange is taken into account.}.
The realistic cross section can be obtained by multiplying
the Born cross section by the corresponding gap survival factor $<S^{2}>$.
In exclusive reactions, as the $p p \to p p \pi^+ \pi^-$ one
for instance, the gap survival factor $<S^{2}>$ is strongly 
dependent on the $t_{1}$ and $t_{2}$ variables, see e.g. 
\cite{Lebiedowicz:2015eka}.
A similar observation was made for the $p p \to p p J/\psi$ reaction \cite{Schafer:2007mm}.
In \cite{Lebiedowicz:2014bea} we have shown that 
the absorption effects due to $pp$-interaction lead to a huge damping
of the cross section for the purely diffractive mechanism 
($<S^{2}> \simeq 0.2$) and a relatively small reduction of 
the cross section for the photoproduction mechanism ($<S^{2}> \simeq 0.9$).
It is not clear whether such a value ($<S^{2}> \simeq 0.9$) 
is relevant for the case of interest (\ref{2to4_reaction}).
Our Born-level cross section, calculated for $\sqrt{s} = 13$~TeV, 
should be compared with $\sigma =$ (206.72 -- 278.80)~nb obtained within
the dipole saturation-inspired approach, 
see Table~1 of \cite{Goncalves:2016uhj}.
Therefore, it seems reasonable to assume that the magnitude of 
the absorptive corrections should be rather larger of order of 
50\%, $<S^{2}> = 0.5$.
We should emphasize that the H1 experimental group
found a similar result for the $\gamma p \to \rho^{0} n \pi^{+}$ reaction \cite{H1:2015bxa}.
From Fig.~\ref{fig:photoprod_rho2}~(c) we find a suppression factor $<S^{2}> = K \approx 0.6 \pm 0.1$.
We leave a detailed analysis of absorption effects for future studies.

\begin{figure}[!ht]
\includegraphics[width=0.44\textwidth]{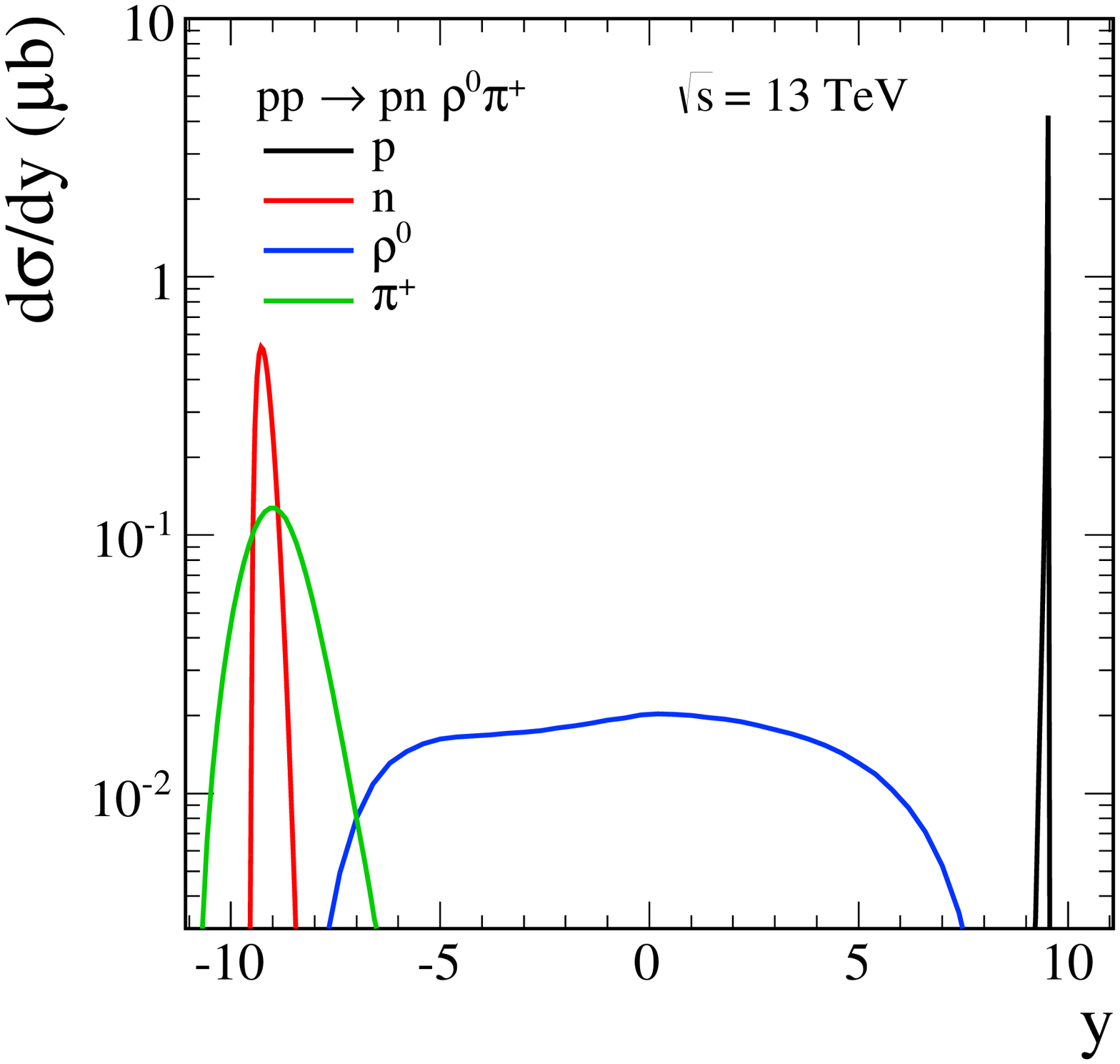}
\includegraphics[width=0.44\textwidth]{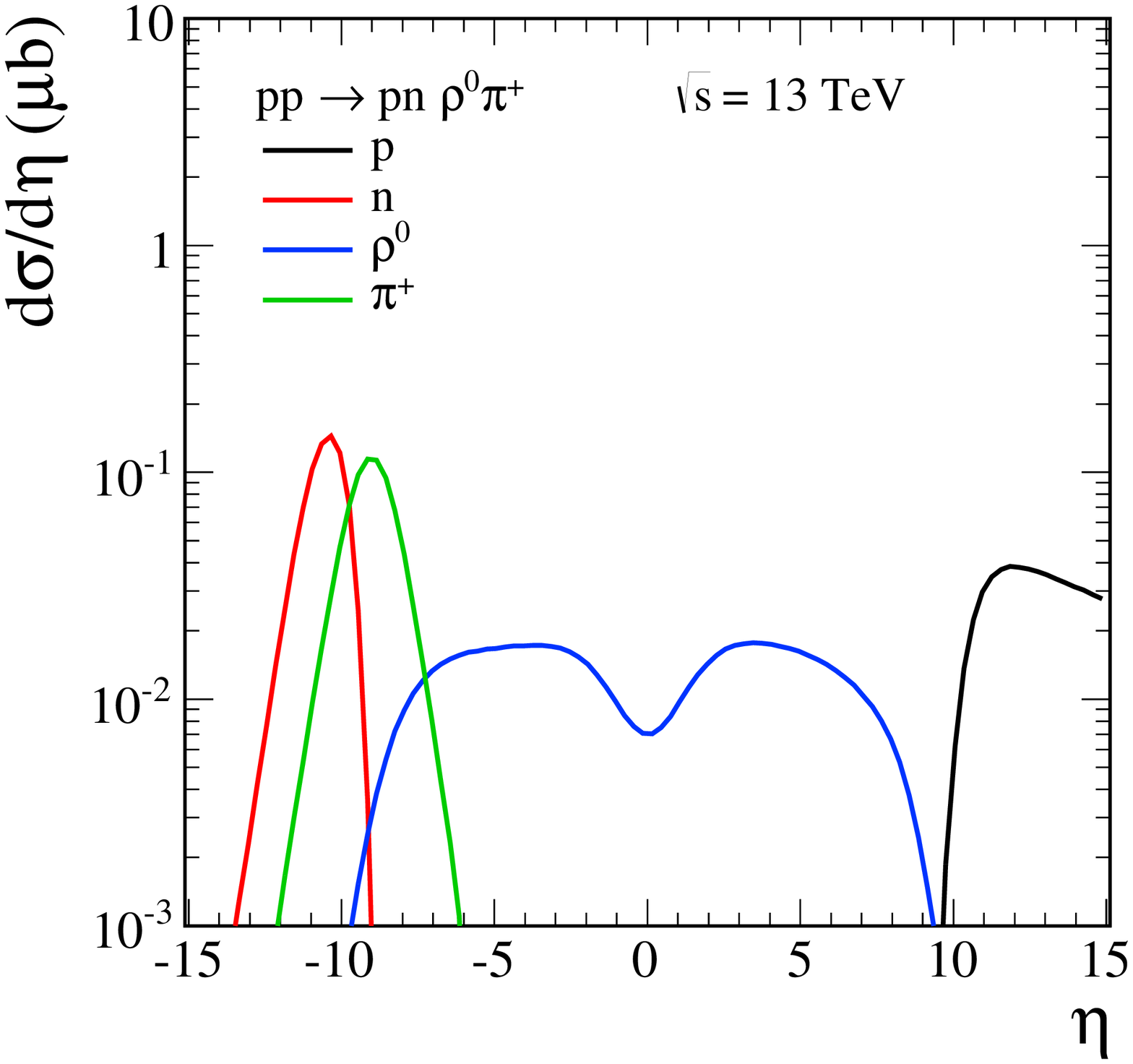}
\includegraphics[width=0.44\textwidth]{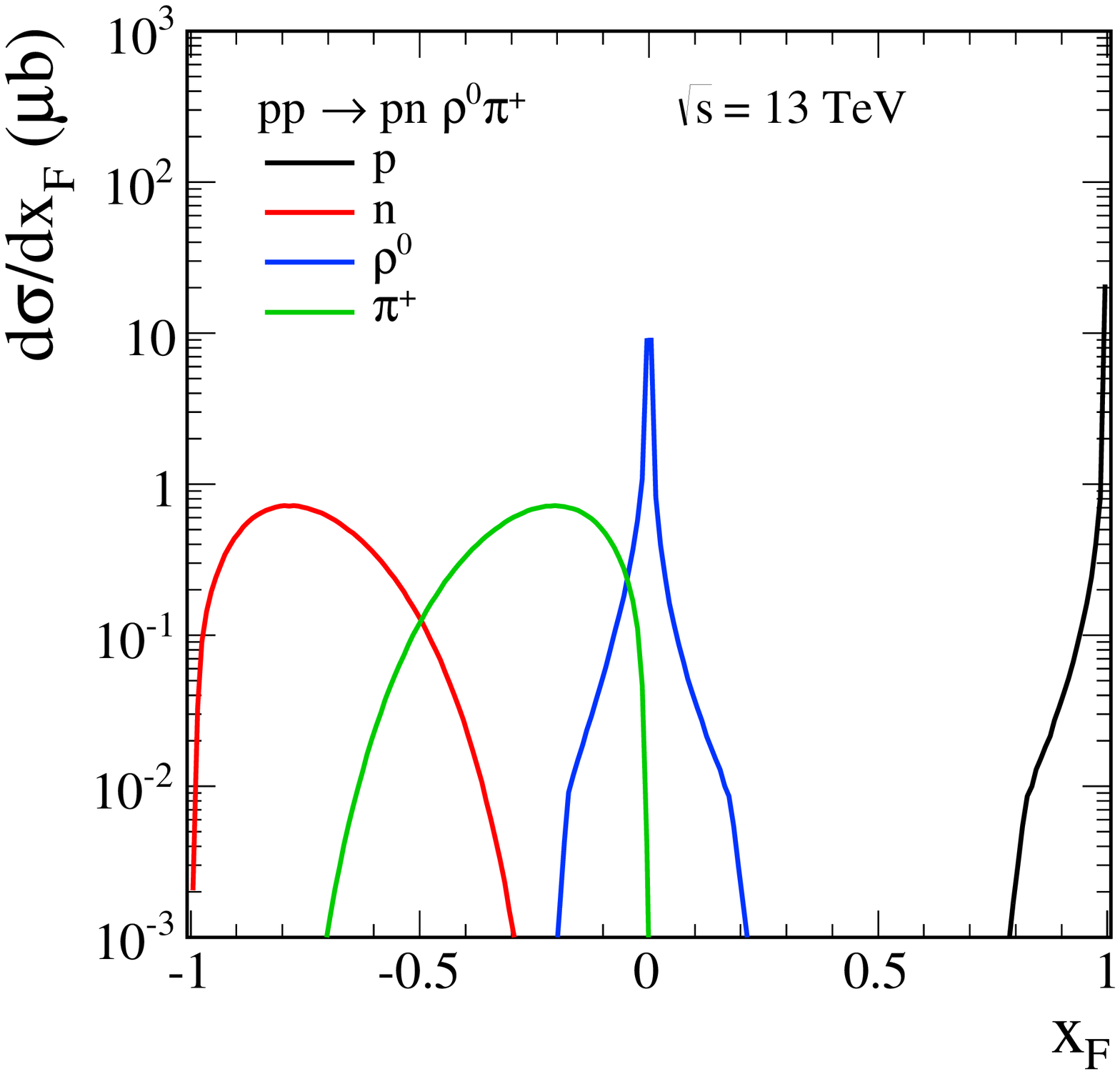}
  \caption{\label{fig:fig1}
  \small
The differential cross sections for the pion exchange mechanism (the diagram of Fig.~\ref{fig:fig0}~(a))
for the reaction $pp \to pn \rho^{0} \pi^{+}$ at $\sqrt{s} = 13$~TeV.
We have taken here $\Lambda = 1$~GeV in (\ref{ff_piNN}).
Absorption effects are not included here.}
\end{figure}

In Fig.~\ref{fig:fig1} we show several distributions
for final state particles (proton, neutron, $\rho^0$ meson
and pion) in several kinematical variables: rapidity, pseudorapidity
and Feynman-$x$ ($x_F = 2 p_{z}/\sqrt{s}$). 
Here we consider only the pion exchange mechanism and only one diagram
where the photon couples to the proton $p (p_{a},\lambda_{a})$; see Fig.~\ref{fig:fig0}~(a).
The second diagram where the r{\^o}les of the two initial protons are interchanged
gives contributions which can be obtained from those
presented here through the replacements 
${\rm y} \to -{\rm y}$, $\eta \to -\eta$ and $x_F \to -x_F$.
The dip in the $\eta$ distribution of $\rho^{0}$ meson
for $|\eta| \to 0$ is a kinematical effect; see Appendix D of \cite{Lebiedowicz:2013ika}.

From the pseudorapidity ($\eta$) distributions 
one can see that it is difficult to perform a fully exclusive measurement.
Different types of detectors must be used:
a central detector (for detection of charged pions from $\rho^{0}$ decay),
very forward proton detectors (ALFA for ATLAS or TOTEM for CMS),
and the Zero Degree Calorimeters (for detection of neutrons).
The $\pi^+$ meson from the reaction (\ref{2to4_reaction}) 
shows up at rapidities $y \approx -11$ to $-6.5$ and is very
difficult to identify with the presently available detectors.

\begin{figure}[!ht]
\includegraphics[width=0.44\textwidth]{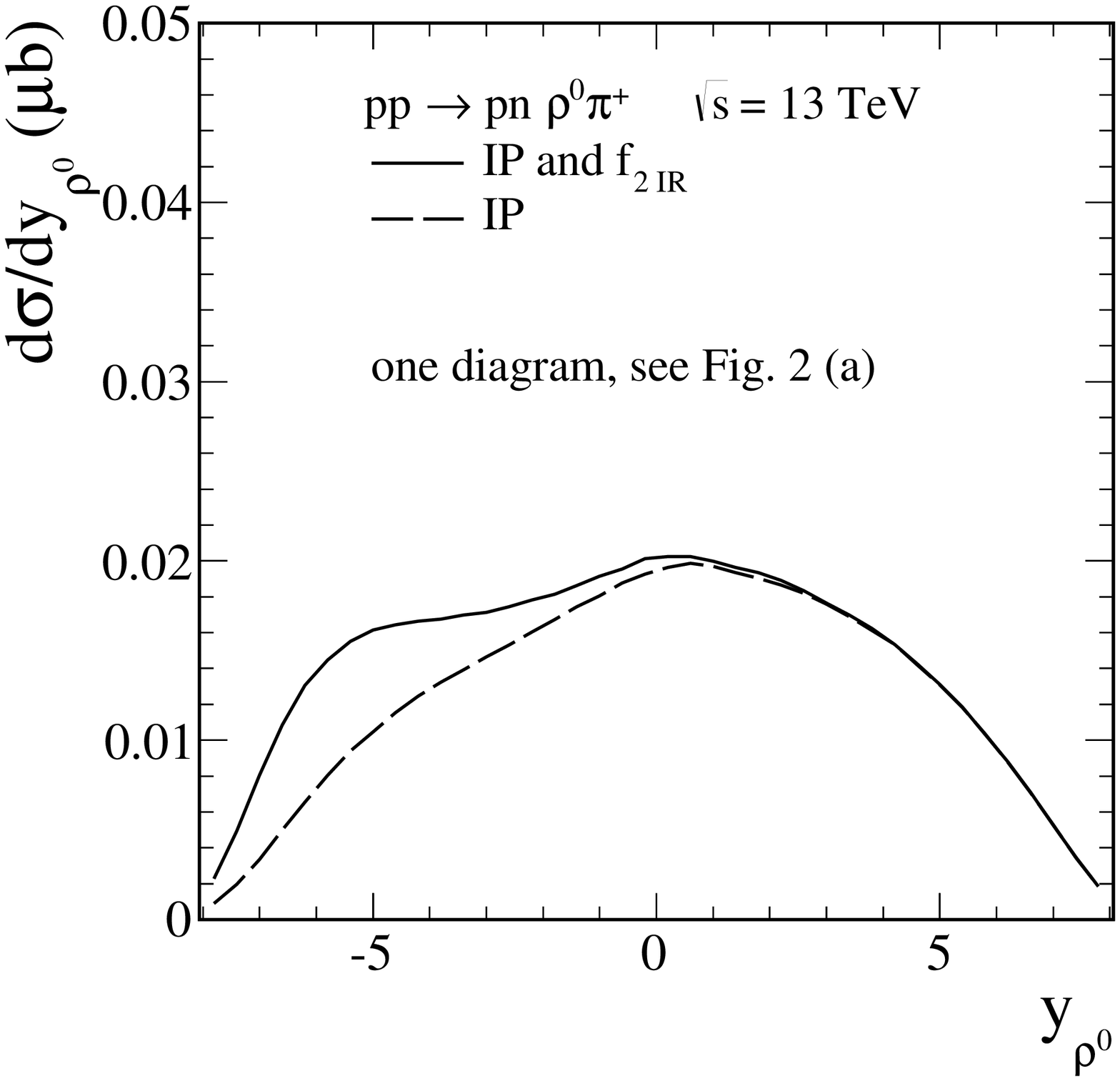}
\includegraphics[width=0.44\textwidth]{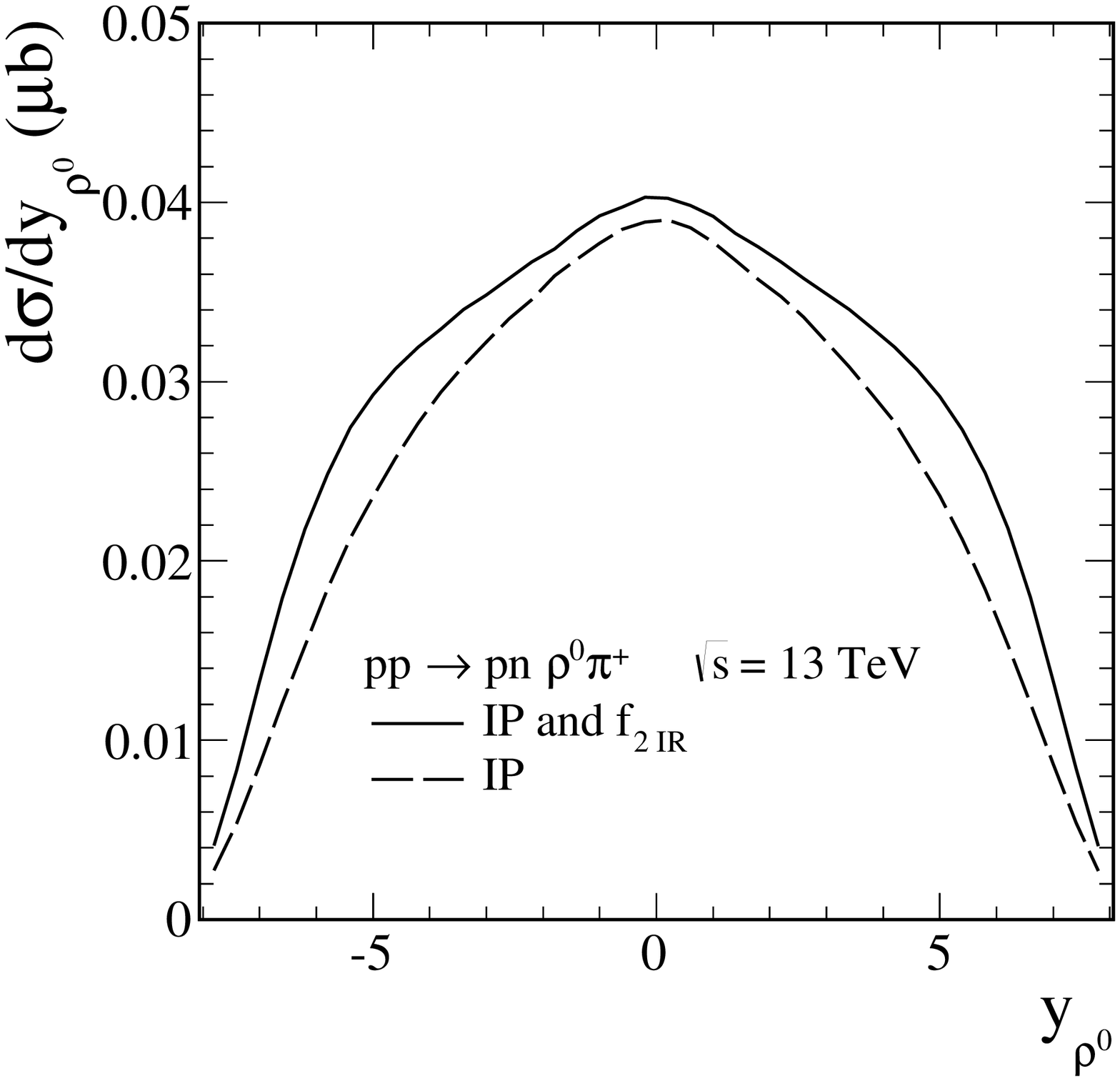}
  \caption{\label{fig:fig1a}
  \small
The distributions in ${\rm y}_{\rho^{0}}$ 
for the reaction $pp \to pn \rho^{0} \pi^{+}$ at $\sqrt{s} = 13$~TeV.
We show the Born-level calculations obtained in the pion exchange mechanism.
In the left panel the results correspond to the diagram
of Fig.~\ref{fig:fig0}~(a)
with tensor pomeron and $f_{2 \Reg}$ exchanges (the solid line) 
and the tensor pomeron exchange contribution alone (the long-dashed line).
In the right panel we show the complete results obtained including two diagrams,
Fig.~\ref{fig:fig0}~(a) plus the one with the r{\^o}le
of the initial protons interchanged.
We have taken here $\Lambda = 1$~GeV in (\ref{ff_piNN}).
}
\end{figure}

In Fig.~\ref{fig:fig1a} we show the rapidity distributions
of $\rho^{0}$ meson produced in the pion exchange mechanism.
The results shown in the left panel are obtained from
the calculation taking only the diagram of Fig.~\ref{fig:fig0}~(a) into account.
In the right panel this diagram
and in addition the diagram where the r{\^o}les of the two initial protons 
are interchanged is taken into account.
The solid line corresponds to the tensor pomeron and $f_{2 \Reg}$ exchanges
while the long-dashed line corresponds to the pomeron exchange alone.
One can observe in the forward/backward rapidity region an enhancement
due to the inclusion of $f_{2 \Reg}$ exchanges.
This may be an interesting point for the LHCb experimental plan in the future
\footnote{The $\pi^{+} \pi^{-}$ pairs from $\rho^{0}$ decay 
could be detected by the LHCb forward spectrometer
which covers the region $2 < \eta < 5$.
There, some experimental limitations on both outgoing pions, 
e.g. a cut on $p_{t,\pi} > 0.2$~GeV, must be imposed.}.

\begin{figure}[!ht]
\includegraphics[width=0.44\textwidth]{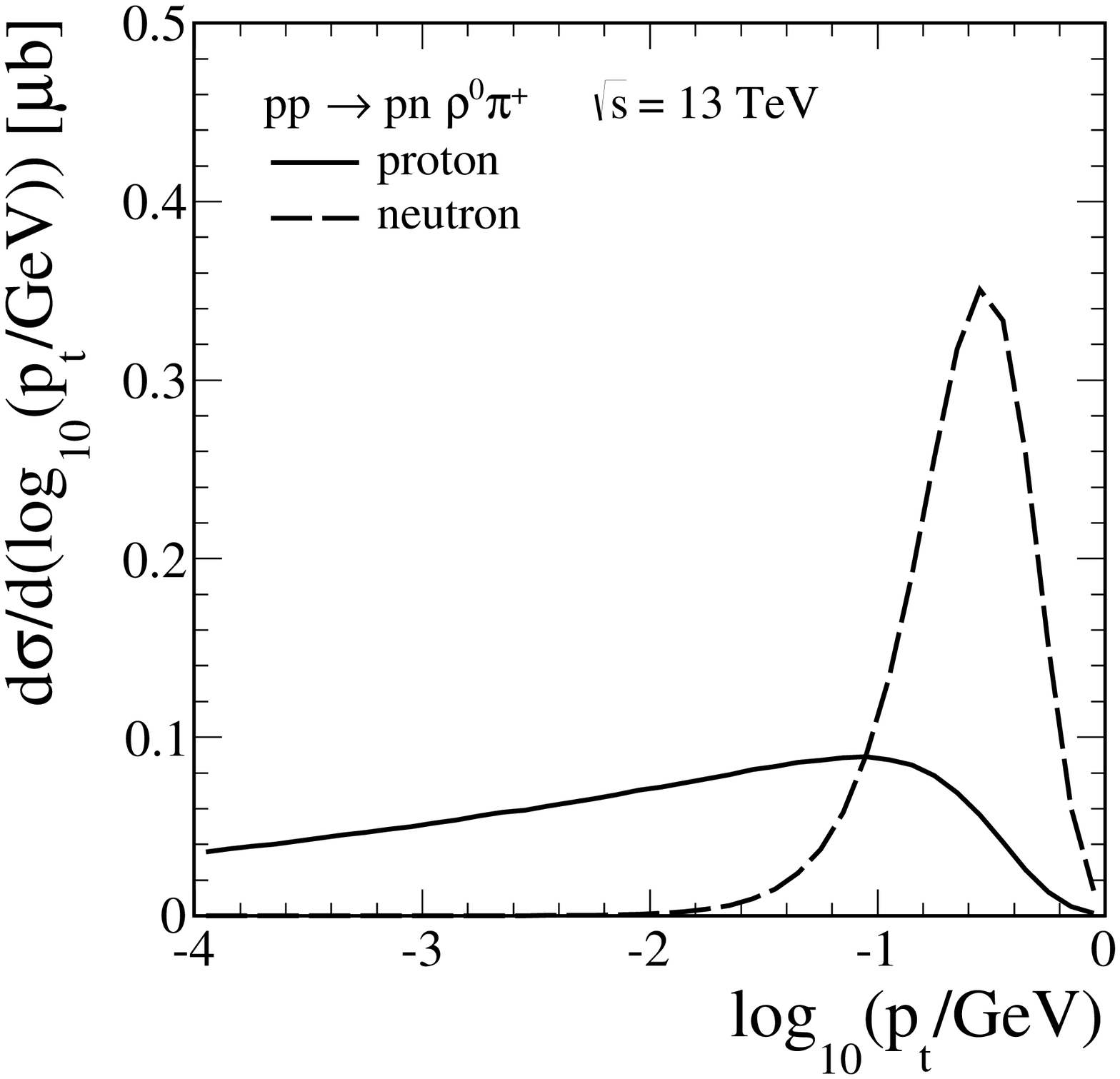}
\includegraphics[width=0.44\textwidth]{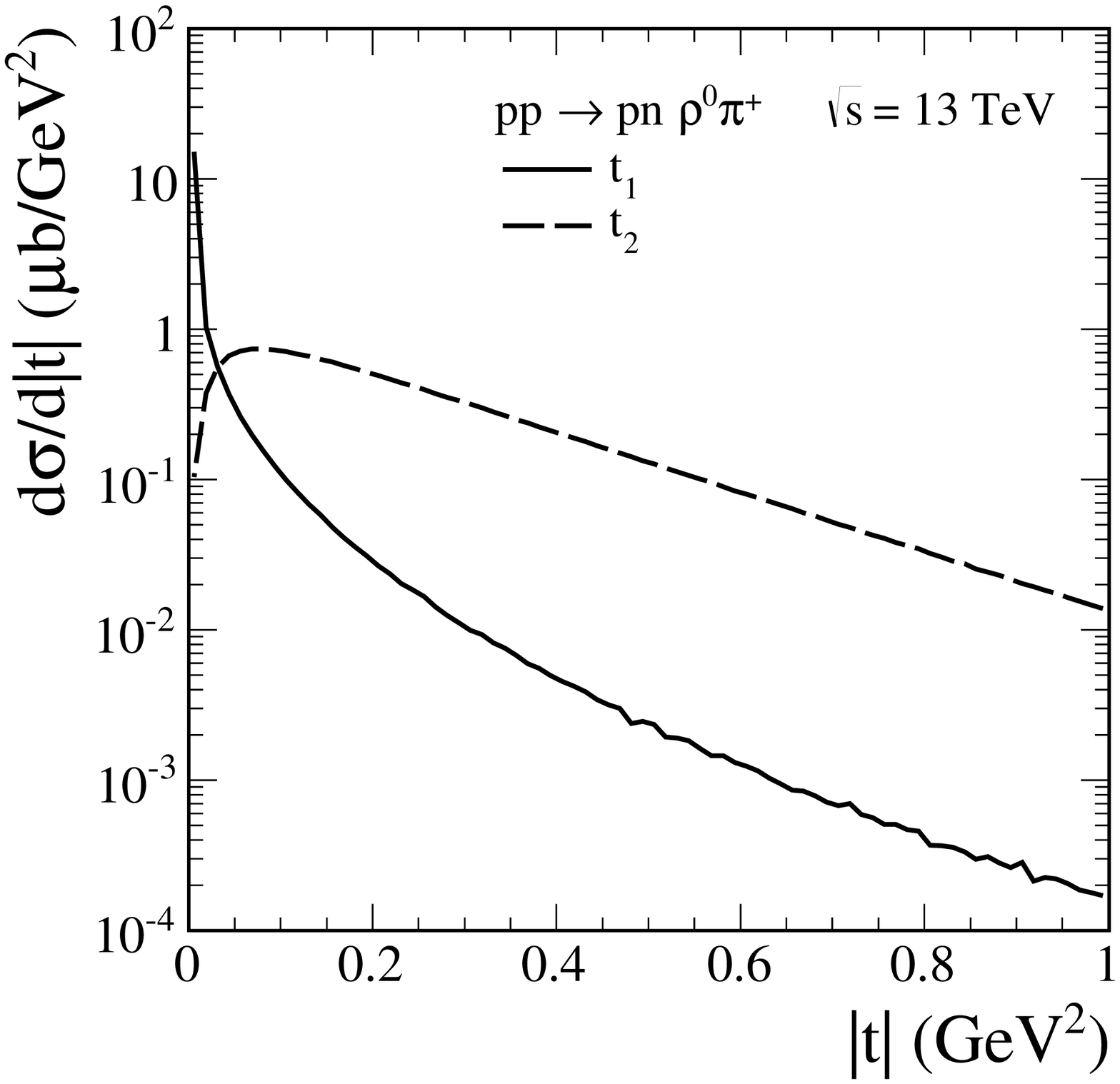}
\includegraphics[width=0.44\textwidth]{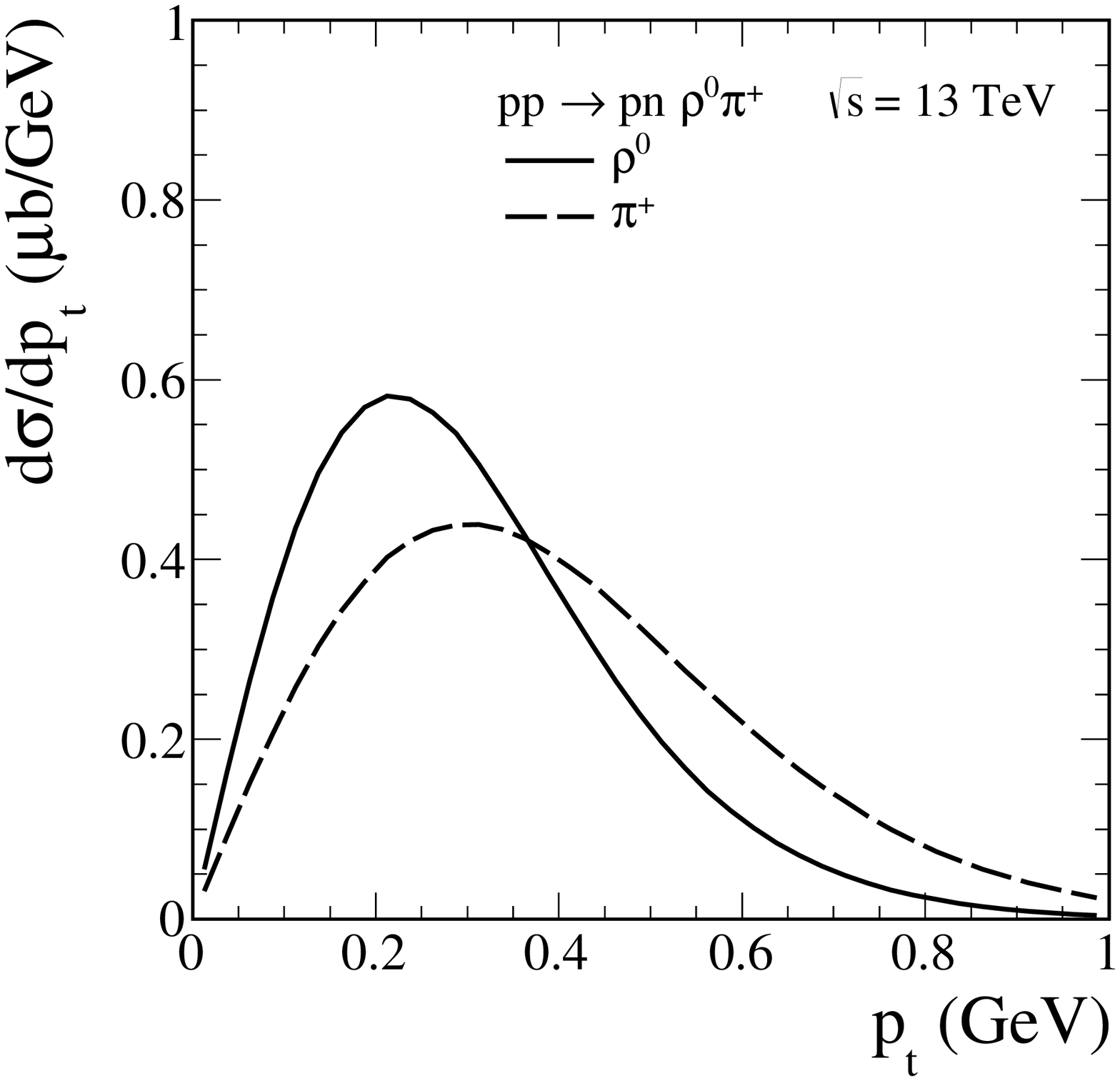}
\includegraphics[width=0.44\textwidth]{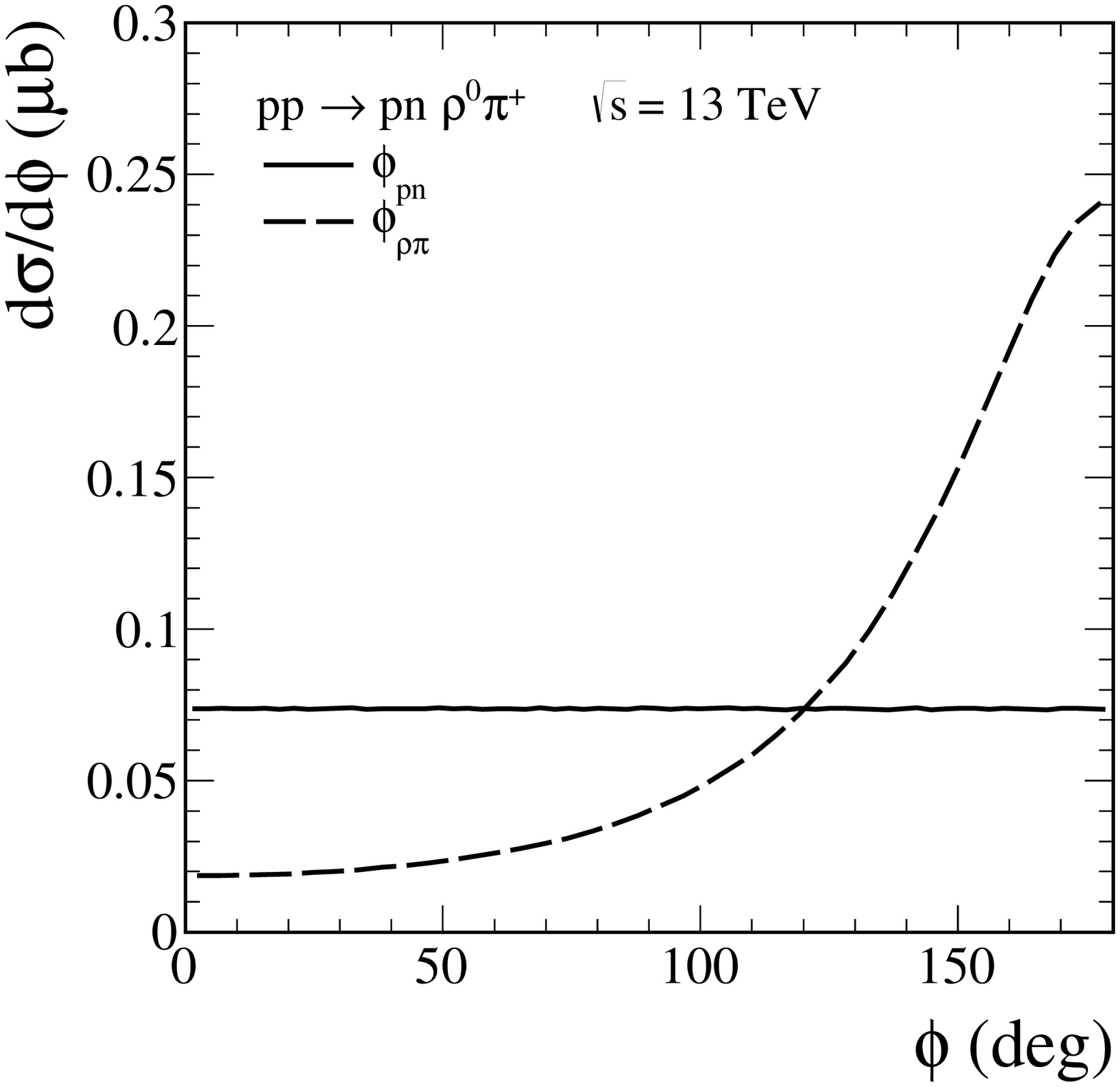}
  \caption{\label{fig:fig2}
  \small
Some differential cross sections for the pion exchange mechanism
for the reaction $pp \to pn \rho^{0} \pi^{+}$ at $\sqrt{s} = 13$~TeV.
Only the diagram of Fig.~\ref{fig:fig0}~(a) is taken into account.
Both tensor pomeron and $f_{2 \Reg}$ exchanges are included in the calculation.
We have taken here $\Lambda = 1$~GeV in (\ref{ff_piNN}).
No absorption effects are included here.}
\end{figure}

Other single particle distributions are shown in Fig.~\ref{fig:fig2}.
In the top-left panel we show the transverse momentum distributions of
proton and neutron. On average protons have much smaller transverse
momenta compared to neutrons. This is easy to understand as due to photon exchange
protons are scattered only at small angles.
In the bottom left panel we show distributions for $\rho^0$
meson and charged pion. Here the differences are much smaller
as the $\rho^0$ meson feels not only photon exchange but also pomeron exchange. 
The four-momentum transfer squared distributions (top right panel) 
are shown for the $p \to p$ vertex (solid line, $t_{1}$) 
and for the $p \to n$ vertex (dashed line, $t_{2}$). 
One can observe a minimum in the $t_{2}$ distribution at $t_{2} = 0$ characteristic for pion exchange.
The correlations in the relative azimuthal angle between proton and neutron
(solid line) and between $\rho^0$ meson and $\pi^+$ (dashed line)
are shown in the bottom-right panel. 
The lack of correlation between proton and neutron can be understood as follows.
The proton $p(p_{a})$ emits a quasi real photon and is scattered to proton $p(p_{1})$;
see (\ref{2to4_reaction}) and Fig.~\ref{fig:fig0}~(a).
The photon travels essentially in the direction of $\vec{p}_{a}$
and its polarisation is determined by the azimuthal angle of $\vec{p}_{1}$.
On the other side of the diagram of Fig.~\ref{fig:fig0}~(a)
the proton $p(p_{b})$ scatters to $n(p_{2})$ emitting a virtual pion $\pi^{+}(q_{2})$.
Thus, in the middle we have the reaction $\gamma \pi^{+} \to \rho^{0} \pi^{+}$.
Without observation of the $\rho^{0}$ polarisation
the cross section for this reaction is independent of the photon polarisation
as follows from parity invariance.
Then, no information from the photon polarisation and thus,
from the azimuthal angle of proton $p(p_{1})$,
can reach the lower part of the diagram,
the $p(p_{b}) \to n(p_{2})$ transition.
Therefore, the azimuthal angles of $\vec{p}_{1}$ and $\vec{p}_{2}$
should be uncorrelated, as indeed we find this from the explicit calculation;
see Fig.~\ref{fig:fig2}, lower right panel.
Also the azimuthal correlation of $\rho^{0}$ and $\pi^{+}$
shown in this figure can be understood from kinematics.
In the $\gamma \pi^{+} \to \rho^{0} \pi^{+}$ reaction
at the center of the diagram of Fig.~\ref{fig:fig0}~(a)
the initial $\gamma$ follows closely the direction of $\vec{p}_{a}$,
the initial $\pi^{+}$ follows, not so closely but still preferentially,
the direction of $\vec{p}_{b}$ (see Fig.~\ref{fig:fig1}, upper right panel).
Then, in the two-body reaction $\gamma \pi^{+} \to \rho^{0} \pi^{+}$
the $\rho^{0}$ and $\pi^{+}$ should come out preferentially at opposite azimuths,
that is, at $\phi = 180^{o}$.
This is indeed what Fig.~\ref{fig:fig2}, lower right panel, shows.

In addition to the diagram of Fig.~\ref{fig:fig0}~(a) there is,
of course, also the diagram with the r{\^o}le of the initial protons interchanged. 
These diagrams contribute to
different corners of the phase space as can be inferred from Fig.~\ref{fig:fig1}.
Thus, there are in essence no interference effects between the amplitudes
from these two diagrams.
This is different for the $p p \to p p V$ processes
where the two amplitudes with the photon coupling to one or the other initial proton
interfere strongly. This leads, for instance,
to an interesting pattern in the distribution of $\phi_{pp}$,
the azimuthal angle between the outgoing protons
(see the discussion in \cite{Schafer:2007mm} 
and also Fig.~13 of \cite{Lebiedowicz:2014bea}).

Finally we compare the cross sections for the inelastic reactions $p p \to p n \rho^0 \pi^+$
and $p p \to p p \rho^0 \pi^0$ to the cross section for the exclusive
elastic process $p p \to p p \rho^0$.
We include here for the inelastic case 
the diagram Fig.~\ref{fig:fig0}~(a)  
and the one with the r{\^o}le of initial protons interchanged and 
also the $p p \rho^0 \pi^0$ production where
the amplitude is as in (\ref{amplitude_approx_2to4}) 
but with the factor $\sqrt{2}$ left out.
For~$\sqrt{s} = 13$~TeV and $\Lambda = 1$~GeV in (\ref{ff_piNN}) we get
\begin{eqnarray}
\sigma_{\rm{inel.}} \equiv \sigma(p p \to p N \rho^0 \pi) = \frac{3}{2} \times 463.2\,{\rm nb} \approx 0.69 \,\mu{\rm b}\,,
\label{inelastic_cross_section_13TeV}
\end{eqnarray}
where $p N \rho^0 \pi$ stands 
for the $p n \rho^0 \pi^{+}$ and $p p \rho^0 \pi^{0}$ final states,
and one of the initial protons, no matter which one, is assumed to dissociate.

For the elastic reaction we get, using the methods of \cite{Lebiedowicz:2014bea},
for~$\sqrt{s} = 13$~TeV
\begin{eqnarray}
\sigma_{\rm{el.}} \equiv \sigma(p p \to p p \rho^0) = 10.32 \,\mu {\rm b}\,.
\label{elastic_cross_section_13TeV}
\end{eqnarray}

For $\sqrt{s} = 7$~TeV the corresponding numbers are
\begin{eqnarray}
\sigma_{\rm{inel.}} = 0.58 \, \mu {\rm b}\,, \quad 
\sigma_{\rm{el.}} = 8.81 \,\mu {\rm b}\,,
\label{cross_sections_7TeV}
\end{eqnarray}
respectively.
For the ratio of inelastic and elastic cross sections we get from
(\ref{inelastic_cross_section_13TeV}) -- 
(\ref{cross_sections_7TeV})
\begin{eqnarray}
&&\frac{\sigma_{\rm{inel.}}}{\sigma_{\rm{el.}}}\mid_{\sqrt{s} = 7\,{\rm TeV}} 
\simeq  6.58 \times 10^{-2}\,,
\label{ratio_cross_section_7TeV}\\
&&\frac{\sigma_{\rm{inel.}}}{\sigma_{\rm{el.}}}\mid_{\sqrt{s} = 13\,{\rm TeV}} 
\simeq  6.69 \times 10^{-2}\,. 
\label{ratio_cross_section_13TeV}
\end{eqnarray}
We see that this ratio increases only slightly with $\sqrt{s}$.

We must mention here that the choice of the form factor parameter
$\Lambda$ in (\ref{ff_piNN})
affects the size of the inelastic cross sections.
For $\Lambda = 1.2$~GeV and $\sqrt{s} = 13$~TeV we get
$\sigma_{\rm{inel.}} = 1.02 \,\mu{\rm b}$,
and $\sigma_{\rm{inel.}}/\sigma_{\rm{el.}} \simeq  9.88 \times 10^{-2}$.
This should be compared to (\ref{inelastic_cross_section_13TeV})
and (\ref{ratio_cross_section_13TeV}), respectively.

We are also interested how the inelastic to elastic ratio depends on rapidity and transverse
momentum of the $\rho^0$ meson. 
Therefore we define
\begin{eqnarray}
R({\rm y}_{\rho^0}) &=& \frac{d \sigma_{p p \to p N \rho^0 \pi}/d{\rm y}_{\rho^0}}{d \sigma_{p p \to p p \rho^0}/d{\rm y}_{\rho^0}}\,,
\label{ratio_y}
\\
R(p_{t, \rho^0}) &=& \frac{d \sigma_{p p \to p N \rho^0 \pi}/dp_{t,\rho^0}}{d \sigma_{p p \to p p \rho^0}/dp_{t,\rho^0}}\,.
\label{ratio_pt}
\end{eqnarray}

\begin{figure}[!ht]
\includegraphics[width=0.44\textwidth]{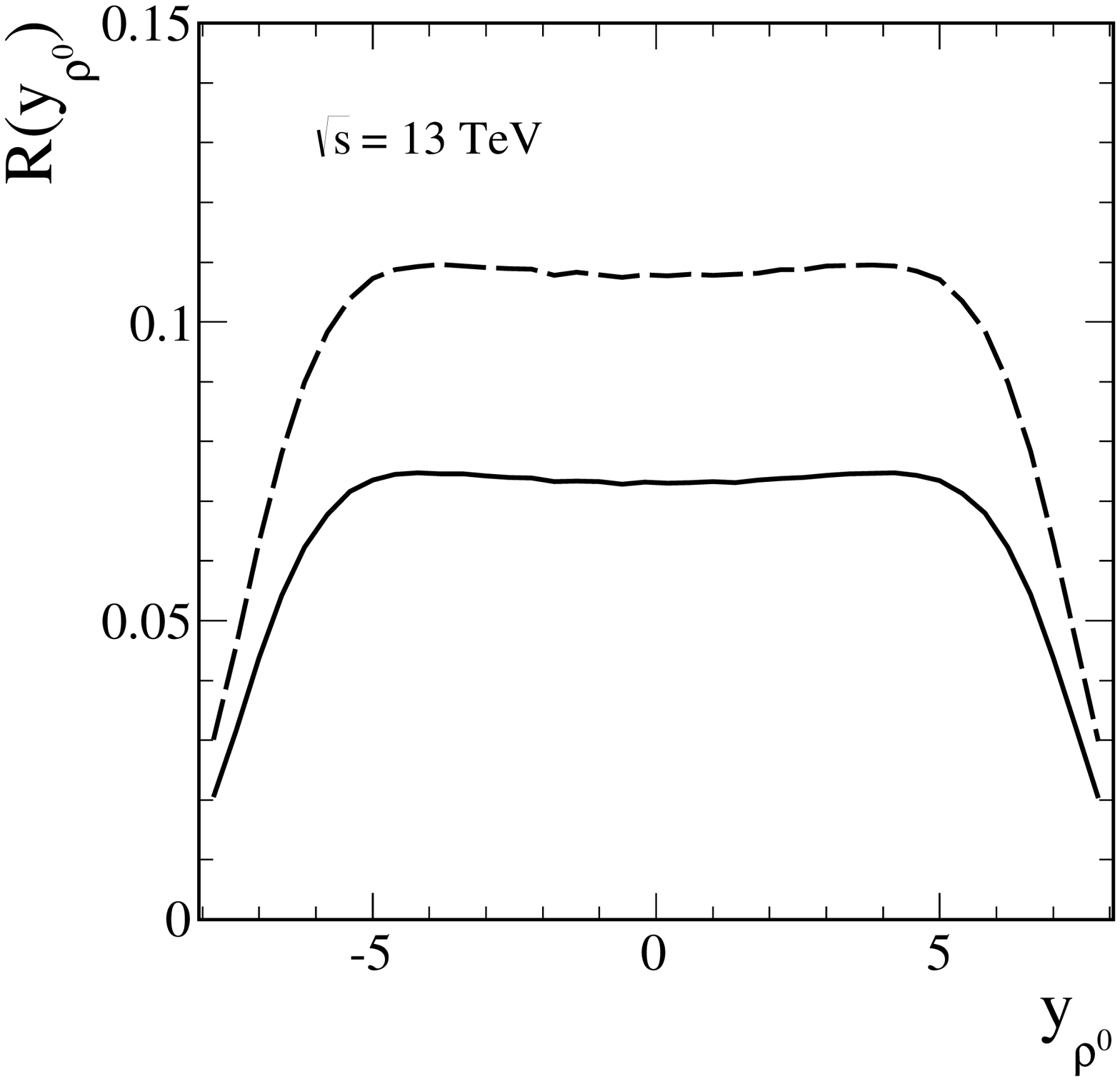}
\includegraphics[width=0.44\textwidth]{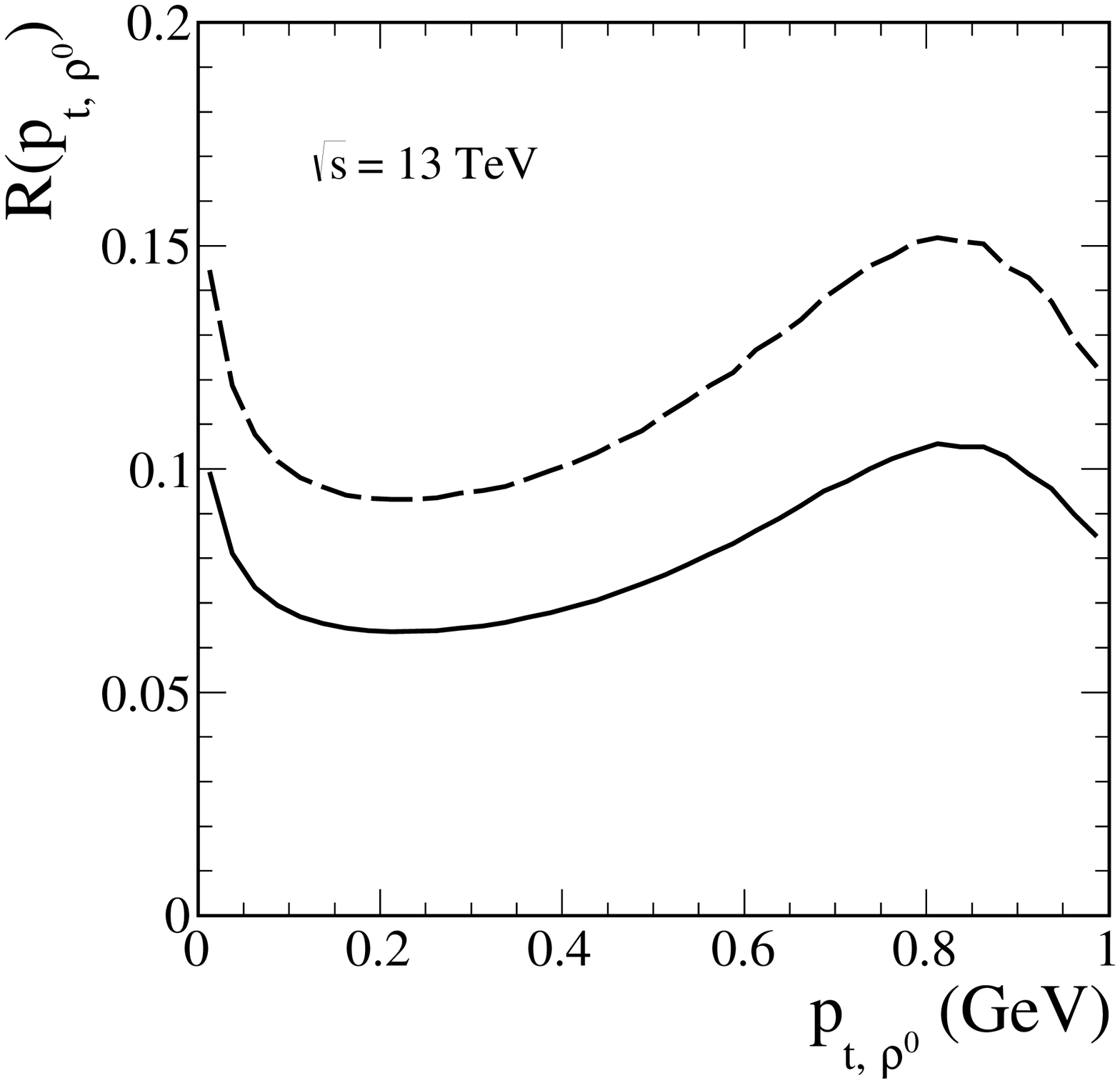}
  \caption{\label{fig:fig3}
  \small
The ratios $R({\rm y}_{\rho^0})$ (\ref{ratio_y}) 
and $R(p_{t,\rho^{0}})$ (\ref{ratio_pt}) at $\sqrt{s} = 13$~TeV.
The solid lines represent results for $\Lambda = 1$~GeV in (\ref{ff_piNN})
and the dashed lines are for $\Lambda = 1.2$~GeV.
No absorption effects are included here
but they should approximately cancel in the ratio.}
\end{figure}

In Fig.~\ref{fig:fig3} we show the ratios as a function of $\rho^0$
rapidity (left panel) and transverse momentum (right panel).
There is almost no dependence of the ratio (\ref{ratio_y}) on rapidity
for midrapidities.
Only at the edges of the rapidity phase space this ratio drops considerably.
For the ratio (\ref{ratio_pt})
we predict an interesting pattern as a function of $p_{t,\rho^{0}}$.

\section{Conclusions}
\label{sec:concl}

In the present paper we have studied the reaction $p p \to p n \rho^0 \pi^+$.
We have considered the diagram of Fig.~\ref{fig:fig0}~(a)
with photon exchange on one side and 
pomeron plus $f_{2 \Reg}$ reggeon exchanges in the middle 
(between $\pi^+$ and $\rho^0$ meson) of the diagram. 
We have included also the diagram with the r{\^o}le of the initial protons interchanged.
Due to the specificity of the reaction the corresponding amplitudes do not
interfere in practice as some of the particles in the final state
are emitted in different hemispheres (exclusively forward or backward) for the two diagrams. 
This is rather useful technically as the integration over the four-particle final state is in the case considered not easy.

We have presented first results at the Born-level. 
Several differential distributions have
been shown explicitly for individual particles in the final state
($p,n,\rho^{0},\pi^{+}$). Compared to the $p p \to p p \rho^0$ reaction
we find no azimuthal angle correlations between the final proton and neutron
for $p p \to p n \rho^0 \pi^+$.
Absorption effects, not considered here, may change the result slightly,
however, we do not expect a large effect because the photon exchange from the proton makes the reaction fairly peripheral.

It is not clear to us whether the reaction $p p \to p n \rho^0 \pi^+$
can be measured in the future at the LHC.
The centrally produced $\rho^{0}$ meson could be identified 
by measuring two charged pions in the main ATLAS and CMS detectors.
The $\pi^{+} \pi^{-}$ pairs can also be detected by the LHCb forward spectrometer.
The forward and backward protons and neutrons could be measured with the help of forward
proton detectors (ALFA for ATLAS or TOTEM for CMS) and
the Zero Degree Calorimeters, respectively. The very forward going $\pi^+$ is
difficult to identify as there are no detectors in the corresponding region of (pseudo)rapidity.

Independent of experimental feasibility the process $p p \to p N \rho^0 \pi$ is interesting
on more general grounds.
It was one of our motivations to study the size of the cross section and of differential
distributions of the $\rho^0$ meson for the ``inelastic'' 
$p p \to p n \rho^0 \pi^+$ compared to the reference ``elastic'' 
$p p \to p p \rho^0$ reaction. In this context we have also included
the $p p \to p p \rho^0 \pi^0$ ``inelastic'' process which gives a two
times smaller contribution due to isospin symmetry of the $p \to \pi N$ vertex.
We have calculated the corresponding ratios as a function of the $\rho^0$ 
rapidity and transverse momentum. 
The ratio of the integrated cross sections is between 7\% and 10\%.
For the ratios of unintegrated cross sections we found a weak dependence
on $\rho^0$ rapidity and an interesting pattern in the $\rho^0$ transverse momentum dependence.

We have shown that the proton excitation processes 
($p p \to p n \rho^0 \pi^{+}$ and $p p \to p p \rho^0 \pi^{0}$) constitute 
an important inelastic (non-exclusive) background to 
the $p p \to p p \rho^0$ reaction
when the final state protons are not measured and only 
rapidity gap conditions are checked experimentally.
The reaction $p p \to p n \rho^0 \pi^{+}$ considered here
may be a prototype for the reaction $p p \to p n J/\psi \pi^+$. 
There, the mass of the $J/\psi$ provides a (somewhat) ``hard'' scale.
Thus, this latter reaction may also be treated in the pQCD dipole approach \cite{Goncalves:2016uhj} 
or in the pQCD $k_{t}$-factorization approach with unintegrated gluon distributions 
as done e.g. in \cite{Cisek:2014ala} for the simpler $p p \to p p J/\psi$ process.

To summarize: in this article we have studied $\rho^{0}$ production in $pp$ collisions
where one proton undergoes diffractive excitation to an $n \pi^{+}$ or $p \pi^{0}$ system.
These processes contribute in experimental studies of exclusive $\rho^{0}$ production
where only large rapidity gaps around the centrally produced $\rho^{0}$
are checked but the forward and backward going protons are not detected.
Recently, experimental results for this kind of processes have been published
by the CDF \cite{Aaltonen:2015uva} and CMS \cite{CMS:2015diy} collaborations.
We found that $\rho^{0}$ production with single diffractive excitation of one proton
(no matter which one)
to $n \pi^{+}$ plus $p \pi^{0}$ constitutes $\approx$~(7--10)\%
of the purely elastic $\rho^{0}$ production at LHC energies.
This should be useful for background estimates to the elastic $\rho^{0}$ reaction.
But we hope that the inelastic $\rho^{0}$ production
will also be studied for its own sake in the future.
Indeed, all our results depend on the $\Pom \rho \rho$ coupling
which determines the cross section for the reaction $\gamma \pi^{+} \to \rho^{0} \pi^{+}$.
Thus, from a measurement of $p p \to p N \rho^0 \pi$ one would be able
to extract the cross section, total and differential, for $\gamma \pi \to \rho^{0} \pi$.
This would be a continuation of the measurements of these quantities at HERA \cite{H1:2015bxa}
at c.m. energies $W_{\gamma \pi} = 13 - 34$~GeV.
We wish to mention that for the LHC at $\sqrt{s} = 13$~TeV 
one could cover a much broader range of $W_{\gamma \pi}$
but the experimental extraction of the $\gamma \pi \to \rho^{0} \pi$ cross sections
is certainly not easy.
Of course, in order to get really reliable results for $\gamma \pi \to \rho^{0} \pi$
in this way it would be mandatory to have control over absorptive corrections in
$p p \to p N \rho^0 \pi$. But this is a rather difficult subject
and is definitely beyond the scope of the present paper.

\acknowledgments
This research was partially supported by the MNiSW Grant No. IP2014~025173 (Iuventus Plus),
the Polish National Science Centre Grant No. DEC-2014/15/B/ST2/02528 (OPUS),
and by the Center for Innovation and Transfer of Natural Sciences 
and Engineering Knowledge in Rzesz\'ow.

\bibliography{refs}

\providecommand{\href}[2]{#2}\begingroup\raggedright\begin{thebibliography}{10}

\bibitem{Aaltonen:2009kg}
T.~Aaltonen {\em et~al.}, (CDF Collaboration), {\em {Observation of Exclusive
  Charmonium Production and $\gamma \gamma \to \mu^+ \mu^-$ in $p\bar{p}$
  Collisions at $\sqrt{s}$ = 1.96 TeV},}
  \href{http://dx.doi.org/10.1103/PhysRevLett.102.242001}{Phys. Rev. Lett.
  {\bfseries 102} (2009) 242001},
\href{http://arxiv.org/abs/0902.1271}{{arXiv:0902.1271 [hep-ex]}}.

\bibitem{Aaij:2013jxj}
R.~Aaij {\em et~al.}, (LHCb Collaboration), {\em {Exclusive $J/\psi$ and
  $\psi$(2S) production in $pp$ collisions at $\sqrt{s} = 7$ TeV},}
  \href{http://dx.doi.org/10.1088/0954-3899/40/4/045001}{J. Phys. {\bfseries
  G40} (2013) 045001},
\href{http://arxiv.org/abs/1301.7084}{{arXiv:1301.7084 [hep-ex]}}.

\bibitem{Aaij:2014iea}
R.~Aaij {\em et~al.}, (LHCb Collaboration), {\em {Updated measurements of
  exclusive $J/\psi$ and $\psi$(2S) production cross-sections in $pp$
  collisions at $\sqrt{s}=7$ TeV},}
  \href{http://dx.doi.org/10.1088/0954-3899/41/5/055002}{J. Phys. {\bfseries
  G41} (2014) 055002},
\href{http://arxiv.org/abs/1401.3288}{{arXiv:1401.3288 [hep-ex]}}.

\bibitem{LHCb:2016oce}
(LHCb Collaboration), {\em {Central exclusive production of $J/\psi$ and
  $\psi(2S)$ mesons in pp collisions at $\sqrt{s}=13$ TeV},}
LHCb-CONF-2016-007, CERN-LHCb-CONF-2016-007.

\bibitem{McNulty:2016sor}
R.~McNulty, {\em {Central Exclusive Production at LHCb},} PoS {\bfseries
  DIS2016} (2016) 181,
\href{http://arxiv.org/abs/1608.08103}{{arXiv:1608.08103 [hep-ex]}}.

\bibitem{Aaij:2015kea}
R.~Aaij {\em et~al.}, (LHCb Collaboration), {\em {Measurement of the exclusive
  $\Upsilon$ production cross-section in $pp$ collisions at $\sqrt{s}=7$ TeV
  and 8 TeV},} \href{http://dx.doi.org/10.1007/JHEP09(2015)084}{JHEP {\bfseries
  09} (2015) 084},
\href{http://arxiv.org/abs/1505.08139}{{arXiv:1505.08139 [hep-ex]}}.

\bibitem{Schafer:2016gwq}
W.~Sch{\"a}fer, A.~Szczurek, and A.~Cisek, {\em {Photoproduction of $J/\psi$
  and $\Upsilon$ states in exclusive and proton-dissociative diffractive
  events},} PoS {\bfseries DIS2016} (2016) 205,
  \href{http://arxiv.org/abs/1607.00900}{{arXiv:1607.00900 [hep-ph]}}.

\bibitem{Cisek:2016kvr}
A.~Cisek, W.~Sch{\"a}fer, and A.~Szczurek, {\em {Semiexclusive production of
  $J/\psi$ mesons in proton-proton collisions with electromagnetic and
  diffractive dissociation of one of the protons},}
\href{http://arxiv.org/abs/1611.08210}{{arXiv:1611.08210 [hep-ph]}}.

\bibitem{ATLAS:2007aa}
P.~Jenni, M.~Nessi, and M.~Nordberg, (ATLAS Collaboration), {\em {Zero Degree
  Calorimeters for ATLAS},}
CERN-LHCC-2007-001, LHCC-I-016.

\bibitem{Grachov:2008qg}
O.~A. Grachov {\em et~al.}, (CMS Collaboration), {\em {Performance of the
  combined zero degree calorimeter for CMS},} CMS-CR-2008-038,
  \href{http://dx.doi.org/10.1088/1742-6596/160/1/012059}{J.Phys.Conf.Ser.
  {\bfseries 160} (2009) 012059},
\href{http://arxiv.org/abs/0807.0785}{{arXiv:0807.0785 [nucl-ex]}}.

\bibitem{Drell:1961zza}
S.~D. Drell and K.~Hiida, {\em {Quasi-Elastic Peak in High-Energy
  Nucleon-Nucleon Scattering},}
\href{http://dx.doi.org/10.1103/PhysRevLett.7.199}{Phys.Rev.Lett. {\bfseries 7}
  (1961) 199--202}.

\bibitem{Deck:1964hm}
R.~T. Deck, {\em {Kinematical interpretation of the first $\pi - \rho$
  resonance},}
\href{http://dx.doi.org/10.1103/PhysRevLett.13.169}{Phys.Rev.Lett. {\bfseries
  13} (1964) 169--173}.

\bibitem{Tsarev:1974nr}
V.~A. Tsarev, {\em {Nucleon diffractive dissociation. I. Peripheral model with
  absorption},}
\href{http://dx.doi.org/10.1103/PhysRevD.11.1864}{Phys.Rev. {\bfseries D11}
  (1975) 1864}.

\bibitem{Berger:1975qa}
E.~L. Berger and P.~Piril{\"a}, {\em {Absorptive effects in exclusive
  diffraction dissociation},}
\href{http://dx.doi.org/10.1103/PhysRevD.12.3448}{Phys.Rev. {\bfseries D12}
  (1975) 3448}.

\bibitem{Ponomarev:1975ru}
L.~A. Ponomarev, {\em {Description of Exclusive Processes in the Regge One Pion
  Exchange Model},}
Sov. J. Part. Nucl. {\bfseries 7} (1976) 70.

\bibitem{Ponomarev:1976nv}
L.~A. Ponomarev, {The description of the low multiplicity reactions}, in {\em
  {Proceedings, 18th International Conference on High Energy Physics, Tbilisi,
  USSR, Jul 15-21, 1976}}, pp.~A1. 24--27.
\newblock 1976.
\newblock
\url{http://inspirehep.net/record/116914/files/c76-07-15-p024.pdf}.
\newblock

\bibitem{Zotov:1978}
N.~P. Zotov and V.~A. Tsarev, {\em {Diffraction dissociation and the
  Drell-Hiida-Deck model},} Sov. J. Part. Nucl. {\bfseries 9} (1978) 266.

\bibitem{Tarasiuk:1979ss}
L.~Tarasiuk, {\em {Absorptive effects in nucleon diffraction dissociation},}
Acta Phys.Polon. {\bfseries B10} (1979) 901--910.

\bibitem{Lebiedowicz:2013vya}
P.~Lebiedowicz and A.~Szczurek, {\em {Exclusive $p p \to p p \pi^{0}$ reaction
  at high energies},} \href{http://dx.doi.org/10.1103/PhysRevD.87.074037}{Phys.
  Rev. {\bfseries D87} (2013) 074037},
\href{http://arxiv.org/abs/1303.2882}{{arXiv:1303.2882 [hep-ph]}}.

\bibitem{Aaron:2010ab}
F.~D. Aaron {\em et~al.}, (H1 Collaboration), {\em {Measurement of Leading
  Neutron Production in Deep-Inelastic Scattering at HERA},}
  \href{http://dx.doi.org/10.1140/epjc/s10052-010-1369-4}{Eur. Phys. J.
  {\bfseries C68} (2010) 381--399},
\href{http://arxiv.org/abs/1001.0532}{{arXiv:1001.0532 [hep-ex]}}.

\bibitem{Andreev:2014zka}
V.~Andreev {\em et~al.}, (H1 Collaboration), {\em {Measurement of Feynman-$x$
  Spectra of Photons and Neutrons in the Very Forward Direction in
  Deep-Inelastic Scattering at HERA},}
  \href{http://dx.doi.org/10.1140/epjc/s10052-014-2915-2}{Eur. Phys. J.
  {\bfseries C74} no.~6, (2014) 2915},
\href{http://arxiv.org/abs/1404.0201}{{arXiv:1404.0201 [hep-ex]}}.

\bibitem{H1:2015bxa}
V.~Andreev {\em et~al.}, (H1 Collaboration), {\em {Exclusive $\rho^0$ meson
  photoproduction with a leading neutron at HERA},}
  \href{http://dx.doi.org/10.1140/epjc/s10052-015-3863-1}{Eur. Phys. J.
  {\bfseries C76} no.~1, (2016) 41},
\href{http://arxiv.org/abs/1508.03176}{{arXiv:1508.03176 [hep-ex]}}.

\bibitem{Goncalves:2015mbf}
V.~P. Goncalves, F.~S. Navarra, and D.~Spiering, {\em {Exclusive processes with
  a leading neutron in $ep$ collisions},}
  \href{http://dx.doi.org/10.1103/PhysRevD.93.054025}{Phys. Rev. {\bfseries
  D93} no.~5, (2016) 054025},
\href{http://arxiv.org/abs/1512.06594}{{arXiv:1512.06594 [hep-ph]}}.

\bibitem{Goncalves:2016uhj}
V.~P. Goncalves, B.~D. Moreira, F.~S. Navarra, and D.~Spiering, {\em {Exclusive
  vector meson production with a leading neutron in photon-hadron interactions
  at hadronic colliders},}
  \href{http://dx.doi.org/10.1103/PhysRevD.94.014009}{Phys. Rev. {\bfseries
  D94} no.~1, (2016) 014009},
\href{http://arxiv.org/abs/1605.08186}{{arXiv:1605.08186 [hep-ph]}}.

\bibitem{Lebiedowicz:2014bea}
P.~Lebiedowicz, O.~Nachtmann, and A.~Szczurek, {\em {$\rho^0$ and
  Drell-S{\"o}ding contributions to central exclusive production of $\pi^+
  \pi^-$ pairs in proton-proton collisions at high energies},}
  \href{http://dx.doi.org/10.1103/PhysRevD.91.074023}{Phys. Rev. {\bfseries
  D91} (2015) 074023},
\href{http://arxiv.org/abs/1412.3677}{{arXiv:1412.3677 [hep-ph]}}.

\bibitem{Ewerz:2013kda}
C.~Ewerz, M.~Maniatis, and O.~Nachtmann, {\em {A Model for Soft High-Energy
  Scattering: Tensor Pomeron and Vector Odderon},}
  \href{http://dx.doi.org/http://dx.doi.org/10.1016/j.aop.2013.12.001}{Annals
  Phys. {\bfseries 342} (2014) 31--77},
\href{http://arxiv.org/abs/1309.3478}{{arXiv:1309.3478 [hep-ph]}}.

\bibitem{Ewerz:2016onn}
C.~Ewerz, P.~Lebiedowicz, O.~Nachtmann, and A.~Szczurek, {\em {Helicity in
  Proton-Proton Elastic Scattering and the Spin Structure of the Pomeron},}
  \href{http://dx.doi.org/10.1016/j.physletb.2016.10.064}{Phys. Lett.
  {\bfseries B763} (2016) 382--387},
\href{http://arxiv.org/abs/1606.08067}{{arXiv:1606.08067 [hep-ph]}}.

\bibitem{Adamczyk:2012kn}
L.~Adamczyk {\em et~al.}, (STAR Collaboration), {\em {Single spin asymmetry
  $A_N$ in polarized proton-proton elastic scattering at $\sqrt{s}=200$ GeV},}
  \href{http://dx.doi.org/10.1016/j.physletb.2013.01.014}{Phys. Lett.
  {\bfseries B719} (2013) 62--69},
\href{http://arxiv.org/abs/1206.1928}{{arXiv:1206.1928 [nucl-ex]}}.

\bibitem{Lebiedowicz:2016ioh}
P.~Lebiedowicz, O.~Nachtmann, and A.~Szczurek, {\em {Central exclusive
  diffractive production of the $\pi^{+}\pi^{-}$ continuum, scalar, and tensor
  resonances in $pp$ and $p \bar{p}$ scattering within the tensor Pomeron
  approach},} \href{http://dx.doi.org/10.1103/PhysRevD.93.054015}{Phys. Rev.
  {\bfseries D93} (2016) 054015},
\href{http://arxiv.org/abs/1601.04537}{{arXiv:1601.04537 [hep-ph]}}.

\bibitem{Cisek:2011vt}
A.~Cisek, P.~Lebiedowicz, W.~Sch{\"a}fer, and A.~Szczurek, {\em {Exclusive
  production of $\omega$ meson in proton-proton collisions at high energies},}
  \href{http://dx.doi.org/10.1103/PhysRevD.83.114004}{Phys.Rev. {\bfseries D83}
  (2011) 114004},
\href{http://arxiv.org/abs/1101.4874}{{arXiv:1101.4874 [hep-ph]}}.

\bibitem{Jenkovszky:2010ym}
L.~L. Jenkovszky, O.~E. Kuprash, J.~W. L{\"a}ms{\"a}, V.~K. Magas, and
  R.~Orava, {\em {Dual-Regge approach to high-energy, low-mass diffraction
  dissociation},} \href{http://dx.doi.org/10.1103/PhysRevD.83.056014}{Phys.Rev.
  {\bfseries D83} (2011) 056014},
\href{http://arxiv.org/abs/1011.0664}{{arXiv:1011.0664 [hep-ph]}}.

\bibitem{Jenkovszky:2012hf}
L.~Jenkovszky, O.~Kuprash, R.~Orava, and A.~Salii, {\em {Low missing mass,
  single- and double diffraction dissociation at the LHC},}
  \href{http://dx.doi.org/10.1134/S1063778814120072}{Phys. Atom. Nucl.
  {\bfseries 77} no.~12, (2014) 1463--1474},
  \href{http://arxiv.org/abs/1211.5841}{{arXiv:1211.5841 [hep-ph]}}.
[Odessa Astron. Pub.25,102(2012)].

\bibitem{Bolz:2014mya}
A.~Bolz, C.~Ewerz, M.~Maniatis, O.~Nachtmann, M.~Sauter, and A.~Sch{\"o}ning,
  {\em {Photoproduction of $\pi^{+} \pi^{-}$ pairs in a model with
  tensor-pomeron and vector-odderon exchange},}
  \href{http://dx.doi.org/10.1007/JHEP01(2015)151}{JHEP {\bfseries 1501} (2015)
  151},
\href{http://arxiv.org/abs/1409.8483}{{arXiv:1409.8483 [hep-ph]}}.

\bibitem{Donnachie:2002en}
A.~Donnachie, H.~G. Dosch, P.~V. Landshoff, and O.~Nachtmann, {\em {Pomeron
  physics and QCD},}
Camb.Monogr.Part.Phys.Nucl.Phys.Cosmol. {\bfseries 19} (2002) 1--347.

\bibitem{Dumbrajs:1983jd}
O.~Dumbrajs, R.~Koch, H.~Pilkuhn, G.~C. Oades, H.~Behrens, J.~J. de~Swart, and
  P.~Kroll, {\em {Compilation of coupling constants and low-energy
  parameters},}
\href{http://dx.doi.org/10.1016/0550-3213(83)90288-2}{Nucl. Phys. {\bfseries
  B216} (1983) 277}.

\bibitem{Ericson:2000md}
T.~E.~O. Ericson, B.~Loiseau, and A.~W. Thomas, {\em {Determination of the
  pion-nucleon coupling constant and scattering lengths},}
  \href{http://dx.doi.org/10.1103/PhysRevC.66.014005}{Phys.Rev. {\bfseries C66}
  (2002) 014005},
\href{http://arxiv.org/abs/hep-ph/0009312}{{arXiv:hep-ph/0009312 [hep-ph]}}.

\bibitem{Lebiedowicz:2015eka}
P.~Lebiedowicz and A.~Szczurek, {\em {Revised model of absorption corrections
  for the $p p \to p p \pi^{+} \pi^{-}$ process},}
  \href{http://dx.doi.org/10.1103/PhysRevD.92.054001}{Phys. Rev. {\bfseries
  D92} (2015) 054001},
\href{http://arxiv.org/abs/1504.07560}{{arXiv:1504.07560 [hep-ph]}}.

\bibitem{Schafer:2007mm}
W.~Sch{\"a}fer and A.~Szczurek, {\em {Exclusive photoproduction of $J/\psi$ in
  proton-proton and proton-antiproton scattering},}
  \href{http://dx.doi.org/10.1103/PhysRevD.76.094014}{Phys.Rev. {\bfseries D76}
  (2007) 094014},
\href{http://arxiv.org/abs/0705.2887}{{arXiv:0705.2887 [hep-ph]}}.

\bibitem{Lebiedowicz:2013ika}
P.~Lebiedowicz, O.~Nachtmann, and A.~Szczurek, {\em {Exclusive central
  diffractive production of scalar and pseudoscalar mesons; tensorial vs.
  vectorial pomeron},}
  \href{http://dx.doi.org/10.1016/j.aop.2014.02.021}{Annals Phys. {\bfseries
  344} (2014) 301},
\href{http://arxiv.org/abs/1309.3913}{{arXiv:1309.3913 [hep-ph]}}.

\bibitem{Cisek:2014ala}
A.~Cisek, W.~Sch{\"a}fer, and A.~Szczurek, {\em {Exclusive photoproduction of
  charmonia in $\gamma p \to V p$ and $p p \to p V p$ reactions within
  $k_t$-factorization approach},}
  \href{http://dx.doi.org/10.1007/JHEP04(2015)159}{JHEP {\bfseries 1504} (2015)
  159},
\href{http://arxiv.org/abs/1405.2253}{{arXiv:1405.2253 [hep-ph]}}.

\bibitem{Aaltonen:2015uva}
T.~A. Aaltonen {\em et~al.}, (CDF Collaboration), {\em {Measurement of central
  exclusive $\pi^+ \pi^-$ production in $p\bar{p}$ collisions at $\sqrt{s} =
  0.9$ and 1.96 TeV at CDF},}
  \href{http://dx.doi.org/10.1103/PhysRevD.91.091101}{Phys. Rev. {\bfseries
  D91} (2015) 091101},
\href{http://arxiv.org/abs/1502.01391}{{arXiv:1502.01391 [hep-ex]}}.

\bibitem{CMS:2015diy}
(CMS Collaboration), {\em {Measurement of exclusive $\pi^{+} \pi^{-}$
  production in proton-proton collisions at $\sqrt{s} = 7~\mathrm{TeV}$},}
  CMS-PAS-FSQ-12-004.

\end{thebibliography}\endgroup

\end{document}